\documentclass [10pt,a4paper]{article}

\usepackage{amsmath}
\usepackage{amssymb}
\usepackage{graphicx}
\usepackage{subfig}
\usepackage{float}
\usepackage{cite}
\usepackage{setspace}

\begin{document}
\centerline{\bf\Large Parity Asymmetry in the CMBR Temperature}
\centerline{\bf\Large Power Spectrum}
\bigskip

\renewcommand{\thefootnote}{\fnsymbol{footnote}}

\centerline{\bf Pavan K. Aluri\footnote{email: aluri@iitk.ac.in}, Pankaj Jain\footnote{email: pkjain@iitk.ac.in}}

\renewcommand{\thefootnote}{\arabic{footnote}}

\bigskip

\centerline{Department of Physics, Indian Institute of Technology, Kanpur
208016, India}

\bigskip

\begin{center}
\begin{minipage}{0.84\textwidth}\begin{spacing}{1}{\small {\bfseries Abstract:}
We study the 
power asymmetry between even and odd multipoles in the multipolar expansion
of CMB temperature data from WMAP, recently reported in the literature.
We introduce an alternate statistic which probes this effect more 
sensitively. We find that the data is highly anomalous
and consistently outside $2\sigma$ significance level 
in the whole multipole range $l=[2,101]$. We examine the possibility that 
this asymmetry may be caused by the foreground cleaning 
procedure or by residual foregrounds. By direct simulations
we rule out this possibility. We also examine several
possible sub-dominant foregrounds, which might lead to such an asymmetry.
However in all cases we are unable to explain the signal seen in data.
We next examine cleaned maps, using procedures other than the one followed
by the WMAP Science team. Specifically we analysed the maps cleaned
by the IPSE procedure, Needlets and the harmonic ILC procedure. In all
these cases we also find a statistically significant signal of power
asymmetry if the power spectrum is estimated from masked sky. However
the significance level is found to be not as high as that in the
case of WMAP best
fit power spectrum.    
Finally, we test for the contribution of low-$l$ multipoles to the observed power
asymmetry. We find that if we eliminate the first six 
multipoles, $l=[2,7]$,
the significance falls below $2\sigma$ CL. Hence we find that the signal 
gets dominant contribution from low-$l$ modes.} 
\end{spacing}\end{minipage}\end{center}

\section{Introduction}
\label{Intro}

The primary aim of WMAP satellite has been to measure full sky CMBR
temperature anisotropies with great precision (Bennett et al., 2003a).
The primary quantity
of interest from these full sky maps is the temperature power spectrum, which is
used to constrain various cosmological parameters (Hinshaw et al., 2003; Spergel et al., 2003; Larson et al., 2011; Komatsu et al., 2011).
It also gave TE cross power spectrum and E-mode power to a good precision (Kogut et al., 2003; Page et al., 2003; Larson et al., 2011; Komatsu et al., 2011).
Since the release of WMAP 1st year data many large scale
anomalies were reported in the data
(Bennett et al., 2003b, Efstathiou, 2003; de Oliveira-Costa et al., 2004;  Eriksen et al., 2004a,b; Ralston \& Jain, 2004; Copi \& Huterer, 2004;
Schwarz et al., 2004; Land \& Magueijo, 2005a; de Oliveira-Costa \& Tegmark, 2006; Copi et al., 2007; Samal et al., 2008;
Copi et al., 2010; Bennett et al., 2011).
Recently, in (Kim \& Naselsky, 2010),  (see also de Oliveira-Costa et al., 2004;
Land \& Magueijo, 2005b; Gurzadyan et al., 2007; Gruppuso et al., 2011; Hansen et al., 2011;
Ben-David, Kovetz \& Itzhaki, 2011),
it was found that the CMB power in odd multipoles is anomalously more than that in the even multipoles.
Some possible sources of such odd modulations to CMB, like signals from inflationary era
and solar system physics, were discussed in (Groeneboom et al., 2010;
Koivisto \& Mota, 2011; Maris et al., 2011).
This power asymmetry between even and odd multipoles is also addressed as \emph{Parity asymmetry}.

The CMBR temperature fluctuations on a sphere are usually expanded in terms of spherical
harmonics, $Y_{lm}$, as
\begin{equation}
 \Delta T(\theta, \phi) = \sum_{l=0}^{\infty}\sum_{m=-l}^{+l} a_{lm} Y_{lm} (\theta, \phi) \,,
\end{equation}
where $a_{lm}$'s are the multipolar expansion co-efficients. The power spectrum
of CMB is defined as
\begin{equation}
 C_l = \frac{1}{2l+1} \sum_{m=-l}^{+l} a_{lm}^* a_{lm} \,.
\end{equation}
We denote $\mathbb{C}_l = l(l+1)C_l/2\pi$. 
Kim \& Naselsky (2010)
defined the following two quantities 
\begin{eqnarray}
 P^{+} &=&  \sum_{l=2}^{l_{max}} \frac{\left( 1 + (-1)^l \right)}{2} \mathbb{C}_l 
\label{eq:statisticP1}
\end{eqnarray}
and
\begin{eqnarray}
 P^{-} &=&  \sum_{l=2}^{l_{max}} \frac{\left( 1 - (-1)^l \right)}{2} \mathbb{C}_l\,.
\label{eq:statisticP2}
\end{eqnarray}
Here $P^+$ and $P^-$ represent the sum of power in the even and odd 
multipoles respectively.
The parity asymmetry statistic was defined as $P^{+}/P^{-}$, which we refer 
to as ``$P(l_{max})$'' for convenience. By comparing the data (Larson et al., 2011) with
 ``pure'' realizations of CMB signal generated based on $\Lambda CDM$ model, they estimated
the $p-$value to be $0.002$. When this statistic was applied on full-sky $C_l$ recovered
from masked maps, using $KQ85$ mask from WMAP's seven year data release, 
the minimum probability or $p-$value
was found to be $p=0.003$. The $l_{max}$ corresponding to this $p-$values 
was found to be 22 from simulations, which is an \emph{a posteriori} choice. 
By accounting for this
posterior choice of $l_{max}=22$, the probability was estimated to have lowered 
to $0.02$.

In the present paper we study this parity asymmetry in considerable detail.
We first consider an alternate statistic to test for parity asymmetry. This
statistic is found to be more sensitive than the one considered in
(Kim \& Naselsky, 2010). We next investigate whether this asymmetry might arise
due to foregrounds, which are not symmetric under parity. Most of the
foreground contamination, however, gets removed in the process of extracting
the primordial power spectrum. Nevertheless the residual foregrounds might
be sufficiently large to cause parity asymmetry. Hence we use simulated
foreground cleaned maps to test the significance of parity asymmetry
in the WMAP CMBR data. We utilize both the ILC and IPSE cleaning procedure
for this purpose. The simulated maps are generated using random realizations
of CMBR and pre-launch Planck Sky Model for foregrounds. We also allow for the possibility
of some unknown foreground components. Another interesting anomaly found in the CMB
data is the ecliptic dipolar modulation of the CMB power, discovered in
(Eriksen et al., 2004a). This signal is also parity asymmetric and hence
one may suspect that there might be a relationship between this and
the signal discovered in (Kim \& Naselsky, 2010). We study the possibility that a 
dipole modulation of temperature anisotropy might lead to the observed
parity asymmetry. Furthermore we examine whether foreground cleaned maps
obtained using alternate procedures such IPSE, Needlet ILC, etc.,
also show the parity asymmetry observed in the WMAP best fit power
spectrum. Finally we determine the contribution of the 
modes at very low $l$ to the observed asymmetry.

This paper is arranged as follows. In the next section, a different statistic to understand
this even-odd power asymmetry is presented.
Then, in section 3 we present our results obtained from mock cleaned data used to test the effect of foreground
residuals.
In section \ref{section4} we explore the effect of unknown influences on the data which might be modulating
the primordial signal to induce the observed power asymmetry. In section \ref{section5} we present our analysis of
parity asymmetry in IPSE cleaned temperature data and the cleaned maps available using other procedures such
as Needlet ILC, etc.
In section \ref{section6}, our results from implementing different
cut at various low$-l$ modes are shown. Finally, we conclude in section 
\ref{section7}. 

\section{An Alternate Statistic}
\label{section2}

In this section we introduce a different statistic to quantify the parity asymmetry. As we shall
see this statistic is a more sensitive probe than the one given in 
Eqs. [\ref{eq:statisticP1}] and [\ref{eq:statisticP2}].
Instead of taking averages of $'l\,'$ even or odd multipoles, we look at local $'l\,'$ power asymmetry.
It is defined as\footnote[1]{While compiling the references to our present work, we learnt that a similar,
but not identical, statistic was used by (Land and Magueijo, 2005b), which was referenced in
the introduction.}
\begin{equation}
 Q(l_{odd}) = \frac{2}{l_{odd} - 1}\sum_{l=3}^{l_{odd}} \frac{\mathbb{C}_{l-1}}{\mathbb{C}_l}\,,
\label{eq:statisticQ}
\end{equation}
where the maximum, $l_{odd}$, is any odd multipole $l \geq 3$ and the summation
is over all odd multipoles upto $l_{odd}$. Thus $Q(l_{odd})$ is a measure of mean
deviation of the ratio of power in an even mulitpole to it's succeeding odd multipole
from one, if it is present in the data. At low$-l$, since $l(l+1)C_l \, \sim$ constant, statistically
we expect our statistic to fluctuate about one, like $P(l_{max})$.

\section{Statistical Significance of parity asymmetry using pure and foreground
cleaned CMBR maps}
\label{section3}

We test the WMAP seven year best fit CMB temperature power spectrum\footnote[2]{Available at http://lambda.gsfc.nasa.gov/}
for anomalous parity asymmetry against both pure realizations of CMBR and simulated cleaned maps. The pure CMB sky maps
are generated as constrained realizations of best fit theoretical CMB power spectrum from
$\Lambda CDM$ model\footnotemark[2].
The \texttt{synfast} facility of freely available
\texttt{HEALPix}\footnote[3]{Available at http://healpix.jpl.nasa.gov/} software (Gorski et al., 2005)
was used to produce full sky pure CMB realizations at $N_{side}=512$ of \texttt{HEALPix}'s sky
pixelization scheme. We then generate
five raw maps corresponding to each frequency channel in which WMAP makes the observations.
We do so by adding the pure CMB maps with synchrotron, thermal dust and free-free emission
templates from the pre-launch Planck Sky Model 
(PSM)\footnote[4]{http://www.planck.fr/heading79.html}
(PLANCK Blue Book, 2005) available in each of the WMAP's frequency bands.
These maps were convolved with appropriate beam transfer functions of K, Ka, Q, V and W
bands of WMAP\footnotemark[2] simulating the raw satellite data. We have also added Gaussian
random noise in each pixel using the mean rms noise levels in each of the WMAP's frequency
channels provided in it's seven year data release (Jarosik et al., 2011). Strong foreground contamination
to the observed cosmic CMB signal are assumed to be due to galactic synchrotron, dust and free-free emissions. Synchrotron
radiation is emitted from relativistic electrons in cosmic rays spiraling into the galactic
magnetic field. When the dust grains in the interstellar medium get heated, they emit radiation
due to vibrational mode transitions in infrared frequency which is the thermal dust emission.
The free-free emission is due to the electron-ion interactions in the ionized medium between
clusters of galaxies. The full sky simulated raw maps thus generated were cleaned using IPSE method
as described below.

\subsection{IPSE cleaning procedure}
Here we briefly outline the cleaning procedure we employ to clean the simulated raw maps. 
The IPSE cleaning procedure (Tegmark, de Oliveira-Costa \& Hamilton, 2003; Saha, Jain \& Souradeep, 2006;
Eriksen et al., 2007a; Saha et al., 2008; Samal et al., 2010) is a minimum
variance optimization method better suited for multi-channel CMB 
observations such as WMAP and PLANCK. 
It exploits the frequency dependence
of astrophysical foregrounds received in various detection channels, enabling us to efficiently
extract the cosmic signal. The method involves linearly combining various multi-channel maps
in multipole space with appropriate weights as
\begin{equation}
 a_{lm}^{clean} = \sum_{i=1}^{n_c} \hat{w}_l^i \frac{a_{lm}^i}{B_l^i} \,,
\end{equation}
where $a_{lm}^{clean}$ is the clean CMB signal extracted from the raw data, $a_{lm}^{i}$,
acquired from measurements in $n_c$ frequency channels by linearly combining them
with appropriate weights $\hat{w}_l^i$. The $B_l^i$ factors are the symmetrized beam transfer functions
in multipole space corresponding to an $i^{th}$ frequency channel.
The weights are computed using the empirical covariance matrix,
\begin{equation}
 \hat{C}_l^{ij} = \frac{1}{2l+1}\sum_{m=-l}^{+l} {a_{lm}^{i*} a_{lm}^j} \,,
\end{equation}
in the formula,
\begin{equation}
 \hat{\textbf{W}_l} = \frac{\textbf{e}_0^T \hat{\textbf{C}}_l^{-1}}{\textbf{e}_0^T \hat{\textbf{C}}_l^{-1} \textbf{e}_0} \,,
\end{equation}
where $\textbf{e}_0 = (1 ... 1)^T$ is a column vector with $n_c$ unit elements and $\hat{\textbf{W}_l}$ is also a
column vector given by $(\hat{w}_l^1 ... \hat{w}_l^{n_c})^T$. The clean power spectrum is then
given by,
\begin{equation}
 \hat{C}_l^{clean} = \frac{1}{\textbf{e}_0^T \hat{\textbf{C}}_l^{-1} \textbf{e}_0} \,.
\end{equation}
Further, taking into account the spatial variation of the foreground power across the
sky, each map is divided into disjoint sky regions and this procedure is applied iteratively
in each of these sky partitions.

Using the IPSE cleaning procedure we generated an
an ensemble of 800 cleaned maps. 
A residual foreground bias correction was implemented on each of the cleaned maps by subtracting
a bias map estimated from these simulations in pixel space (Bennett et al., 2003c). The power spectrum of each of these
simulated maps was computed using the \texttt{anafast} facility of \texttt{HEALPix}
and corrected for beam and pixel window effects. We used full sky cleaned maps' power spectrum for the range
$l=[2,10]$. The low$-l$ power can be recovered reliably
from full sky cleaned maps using the IPSE method (Tegmark, de Oliveira-Costa \& Hamilton, 2003).
For $l \geq 11$, we used full sky $C_l$ recovered from partial sky map we get after applying
a galactic mask excluding the heavily contaminated regions in the sky.
The $KQ85$ mask\footnotemark[2] provided by WMAP science team in their seven year data release was applied on the
simulated maps and we recovered the full-sky $C_l$ using the MASTER of CMBR
or pseudo$-C_l$ estimator (Hivon et al., 2002).
At low$-l$, up to $l=32$,
the WMAP best fit power spectrum is estimated 
from a low resolution ILC map 
by using the Blackwell-Rao likelihood estimator (Larson et al., 2011).
For $l > 32$, they estimated the multipole power using the same pseudo$-C_l$
estimator that we use to estimate $C_l$ at high$-l$. 
These power spectra form the basis of our analysis.

\subsection{Statistical significance}

\begin{figure}[h]
\centering
\includegraphics[angle=-90,width=0.84\textwidth]{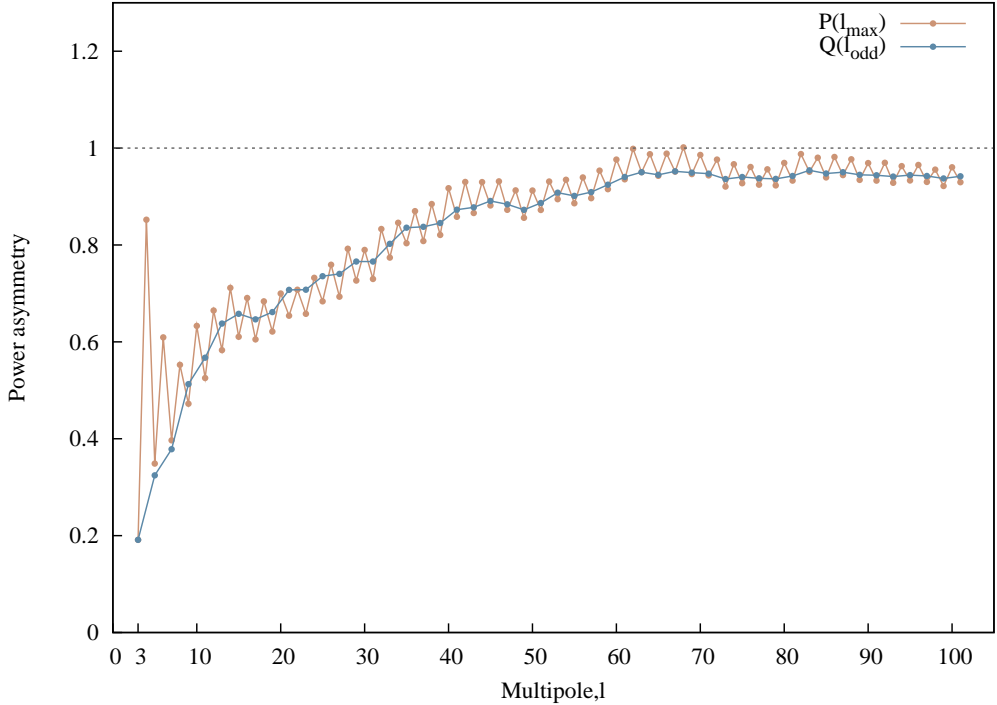}
\caption{The even-odd multipole power asymmetry in WMAP's seven year best fit temperature
          power spectrum in the multipole range $l=[2,101]$ is shown. The asymmetry
          is computed using both the $P(l_{max})$ statistic (lighter curve) and our parity
          asymmetry statistic, $Q(l_{odd})$ (darker curve).}
 \label{KN_OS_bestfit}
\end{figure}

We next compute the statistical significance using both the statistics,
$P(l_{max})$ and $Q(l_{odd})$,
given in Eqs. [\ref{eq:statisticP1}, \ref{eq:statisticP2}] and Eq. [\ref{eq:statisticQ}] respectively.
We used the best fit CMB temperature power spectrum from WMAP's seven
year data release as reference data. The values of both the statistics
for the best fit power are 
shown in Fig. [\ref{KN_OS_bestfit}].

\begin{figure}[h]
 \centering
 \includegraphics[angle=-90,width=0.84\textwidth]{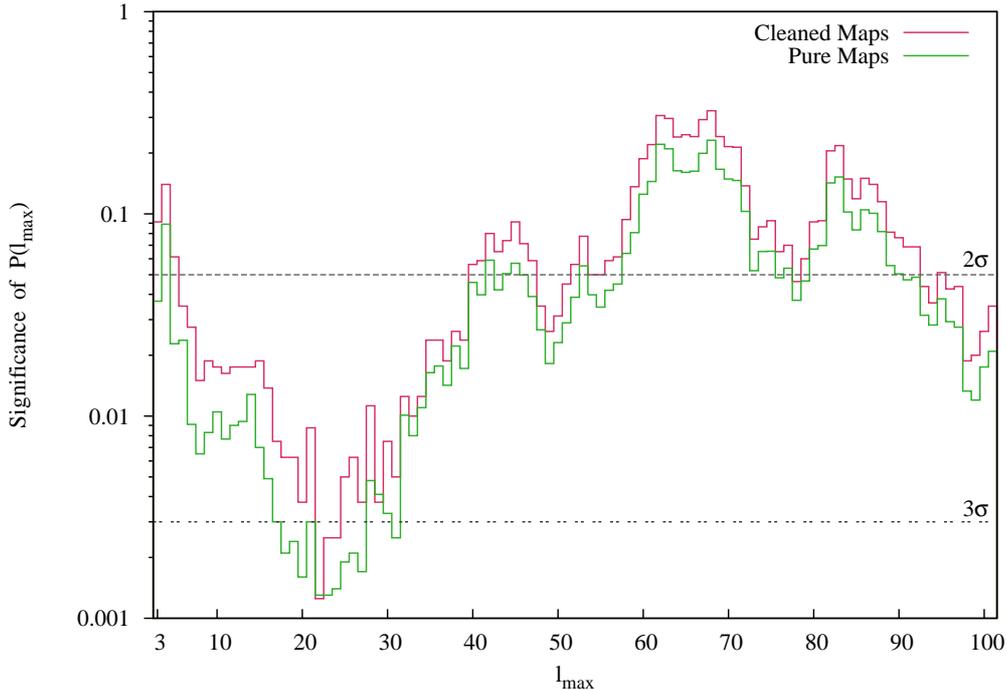}
 \caption{Probability estimates of parity asymmetry seen in the data using the $P(l_{max})$
          parity statistic, in the multipole range $l = [2,101]$. The significances are
          computed using 10,000 pure maps and 800 cleaned maps, cleaned using IPSE method. We find no significant difference between the two estimates.
          As can be seen, the most significant results
          occurs at $l=22$ for both the cases and is beyond $3\sigma$ CL.}
 \label{Signf_KNstat}
\end{figure}

The random chance occurrence probability of getting a $P(l_{max})$ lower than that of the data
for $l_{max} \in [3,101]$ is shown in Fig. [\ref{Signf_KNstat}].
For the case of pure maps we used an ensemble of 10,000 simulated CMBR maps.
We reproduce the results from (Kim \& Naselsky, 2010)
using the pure maps ensemble with lowest probability at $l_{max}=22$.
From the 10,000 pure maps we generated, the probability was found to 
be $0.0013$.
Also, presented in the graph are significances computed using the cleaned maps 
ensemble. In this case
we used only 800 simulated maps due to constraints on computational time. 
As can be seen, the $p-$values from cleaned maps
are slightly higher than the probability estimates from pure maps, but are 
relatively close.
The foreground power have strong even parity preference and, in (Kim \& Naselsky, 2010), the authors
speculated that the observed asymmetry could be due to over subtraction of foregrounds during
foreground reduction. But, eventually they ruled out this possibility. 
Using cleaned maps,
we confirm that this is indeed the case. Thus,
any residual foreground contamination present in the cleaned maps may not induce a
particular parity preference in the data. The contribution of noise is negligible to the power
at low$-l$. Since, we studied this power asymmetry in a wider multipole range, up to $l=101$, and
incorporated noise in the simulated raw maps, we 
also conclude that noise cannot cause
this asymmetry. From the ensemble of cleaned maps, the lowest probability for this parity asymmetry
is again found to be at $l_{max}=22$ with a chance probability of $0.13\%$. This minimum
value for significance is beyond $3\sigma$ CL and quickly falls below $2\sigma$ CL by around $l_{max}=40$. Beyond $l_{max}=40$
it largely stays below $2\sigma$. One interesting thing to note is that the $P(l_{max})$ curve in
Fig. [\ref{KN_OS_bestfit}] looks wavy, like it was overlayed by some oscillations. It may be indicative of
the presence of some underlying modulation (see for example Turner, 1983; Martin \& Ringeval 2004, 2006;
Wang et al., 2005; Ichiki, Nagata \& Yokoyama, 2010).

\begin{figure}[h]
 \centering
 \includegraphics[angle=-90,width=0.84\textwidth]{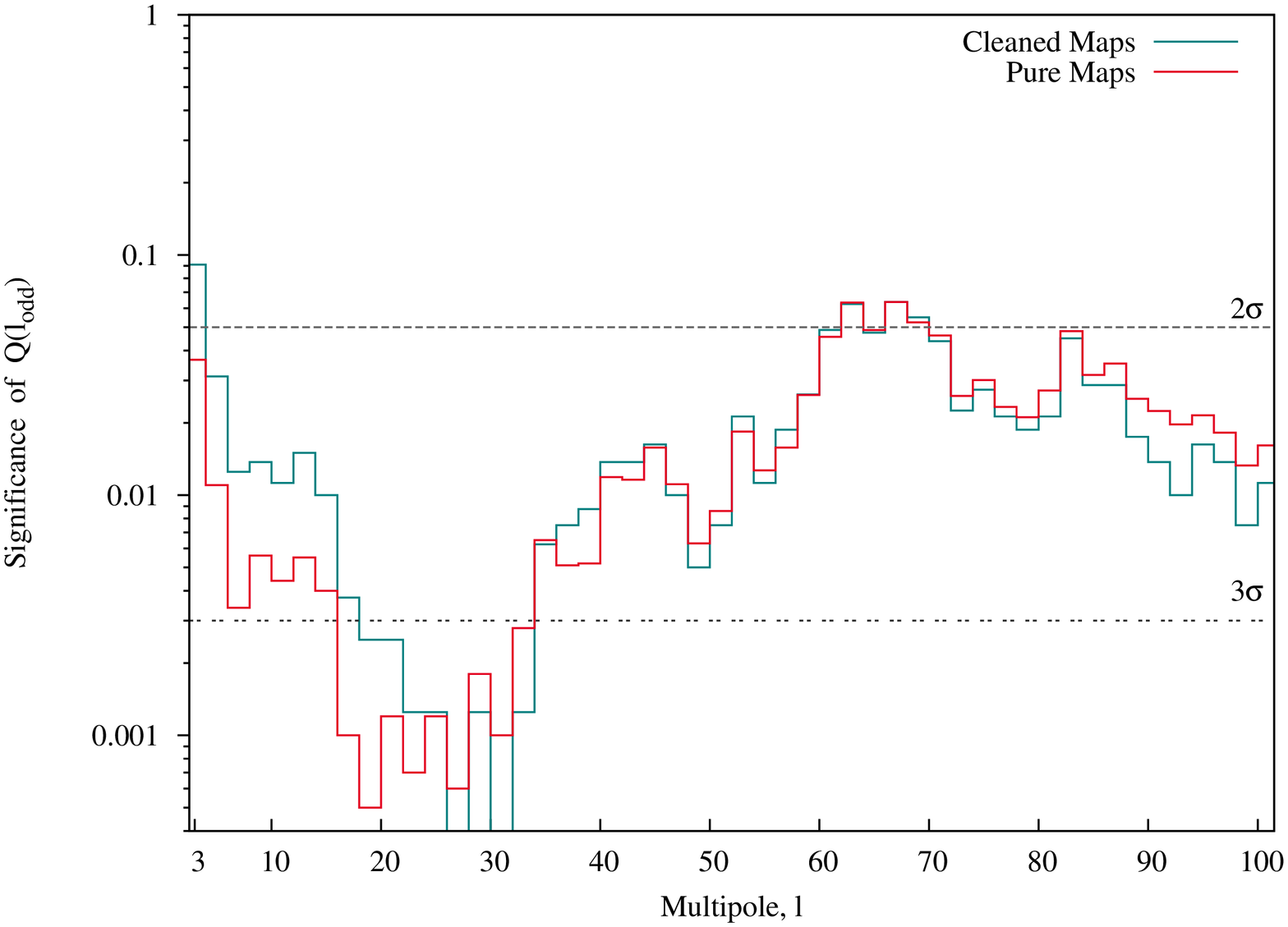}
 \caption{The $p-$values of $Q(l_{odd})$ for WMAP 7 year best fit power spectrum are shown here. As can be seen, the data
          is consistently anomalous by being outside $2\sigma$ in the whole multipole range $l=[2,101]$. For the case of cleaned maps, a value less than 
$1/800$ indicates that we obtained no simulated maps whose statistic was smaller
than 
that of observed data. }
 \label{Signf_ourstat}
\end{figure}

With our estimator $Q(l_{odd})$, we find that, in almost 
the entire multipole range $l=[2,101]$, the
significance of the parity asymmetry lies consistently outside $2\sigma$ CL in both cases using the pure maps and
the cleaned maps. These probability estimates are shown in Fig. [\ref{Signf_ourstat}].
The only exceptions are the significances of the multipoles
63, 67 and 69 with the cleaned maps ensemble 
which are marginally inside the $95\%$ CL. 
Since, $P(l_{max})$ involves the sum of all even or odd multipole power up to a chosen $l_{max}$, which
is equivalent to the mean power up to that $l_{max}$, it appears 
that their sum is hiding the
true significance. As can be seen from the plot, we find that for $l \in [18,31]$ the significance of
parity asymmetry using $Q(l)$ on WMAP's seven year best fit temperature $C_l$ is outside
$3\sigma$ as estimated from pure maps with minimum at $l=19$.  
In  (Kim \& Naselsky, 2010), $l_{max}=22$ is specially singled out, for the $p-$value is lowest at that $l_{max}$. Here, we see
that the $p-$values of $Q(l)$ in the range $l=[18,31]$ remain close to their
minimum. Hence we don't attribute any special
significance to a particular multipole where $Q(l)$ is minimum, but rather to the whole range $l=[18,31]$. With the
cleaned maps simulated set, we find that the $p-$value curve has slightly
rised, but only slightly in comparison to that of pure maps. Hence
we argue that residuals in the cleaned data cannot cause the observed power asymmetry between even and odd multipoles.
In the case of cleaned maps, the minimum probability for this parity asymmetry is found in the range $l=[22,33]$ using our statistic.

In our analysis above we used the IPSE cleaning procedure on the simulated
maps. It is clearly better to use the same procedure for cleaning both the
observed data as well as the simulated maps. We do this in section 5.1 where
we use the ILC cleaning procedure uniformly for the entire analysis. As we
shall see the results in that case are also consistent with those obtained
in the present section.

\section{Unknown foregrounds}
\label{section4}

\begin{figure}[p]
  \centering
  \subfloat{\includegraphics[width=0.7\textwidth]{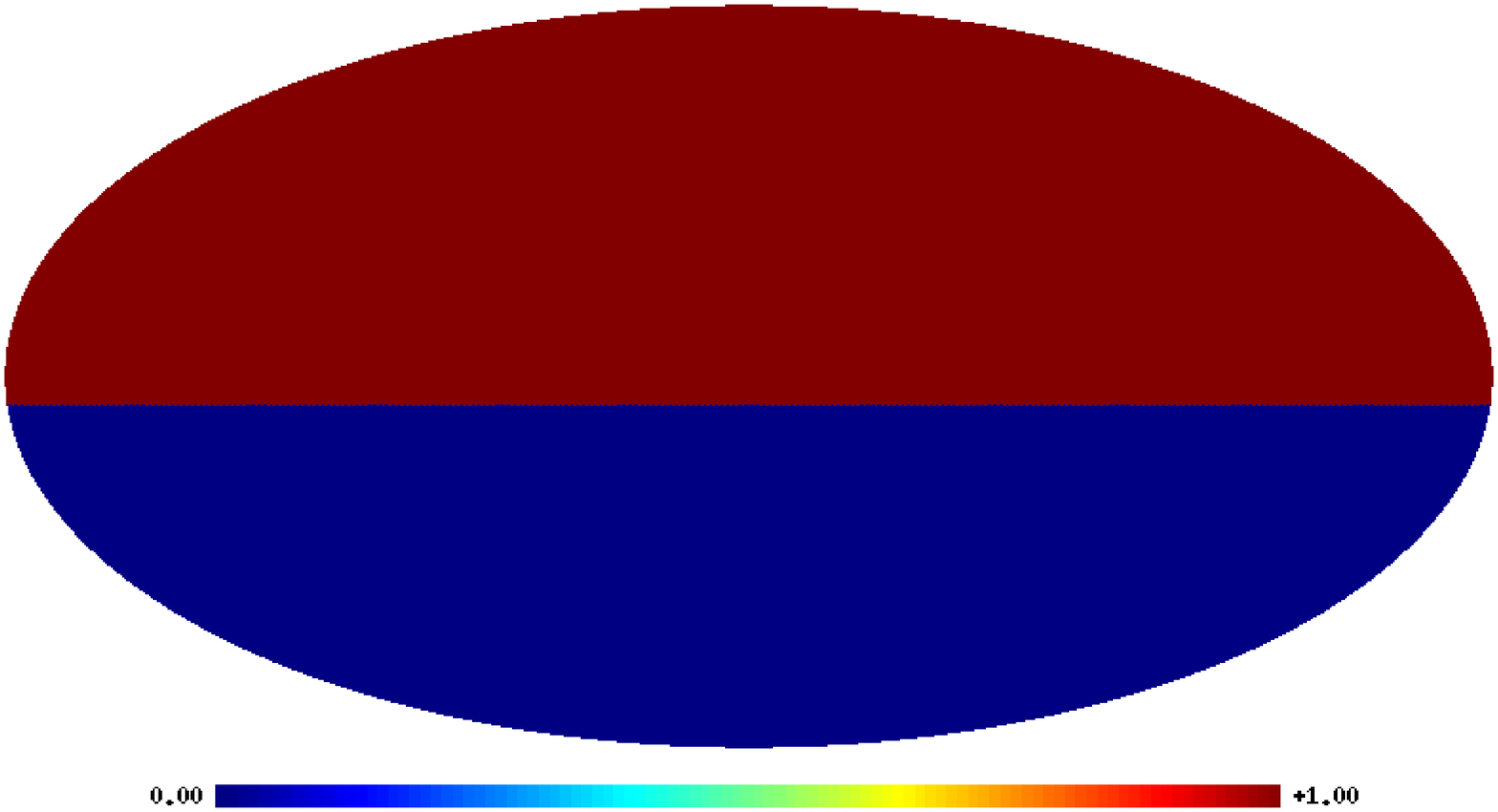}} \\
  \subfloat{\includegraphics[width=0.7\textwidth]{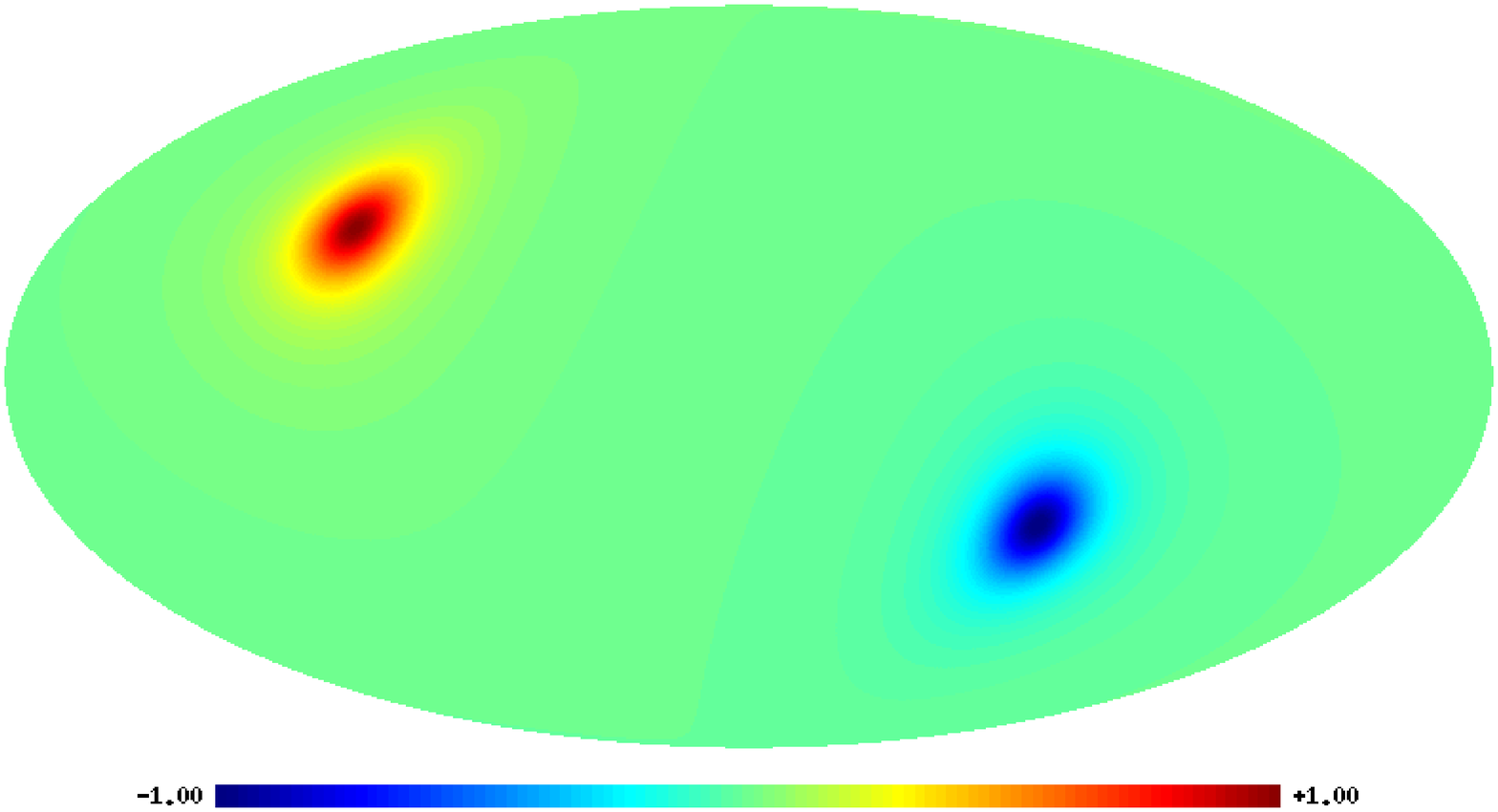}} \\
  \subfloat{\includegraphics[width=0.7\textwidth]{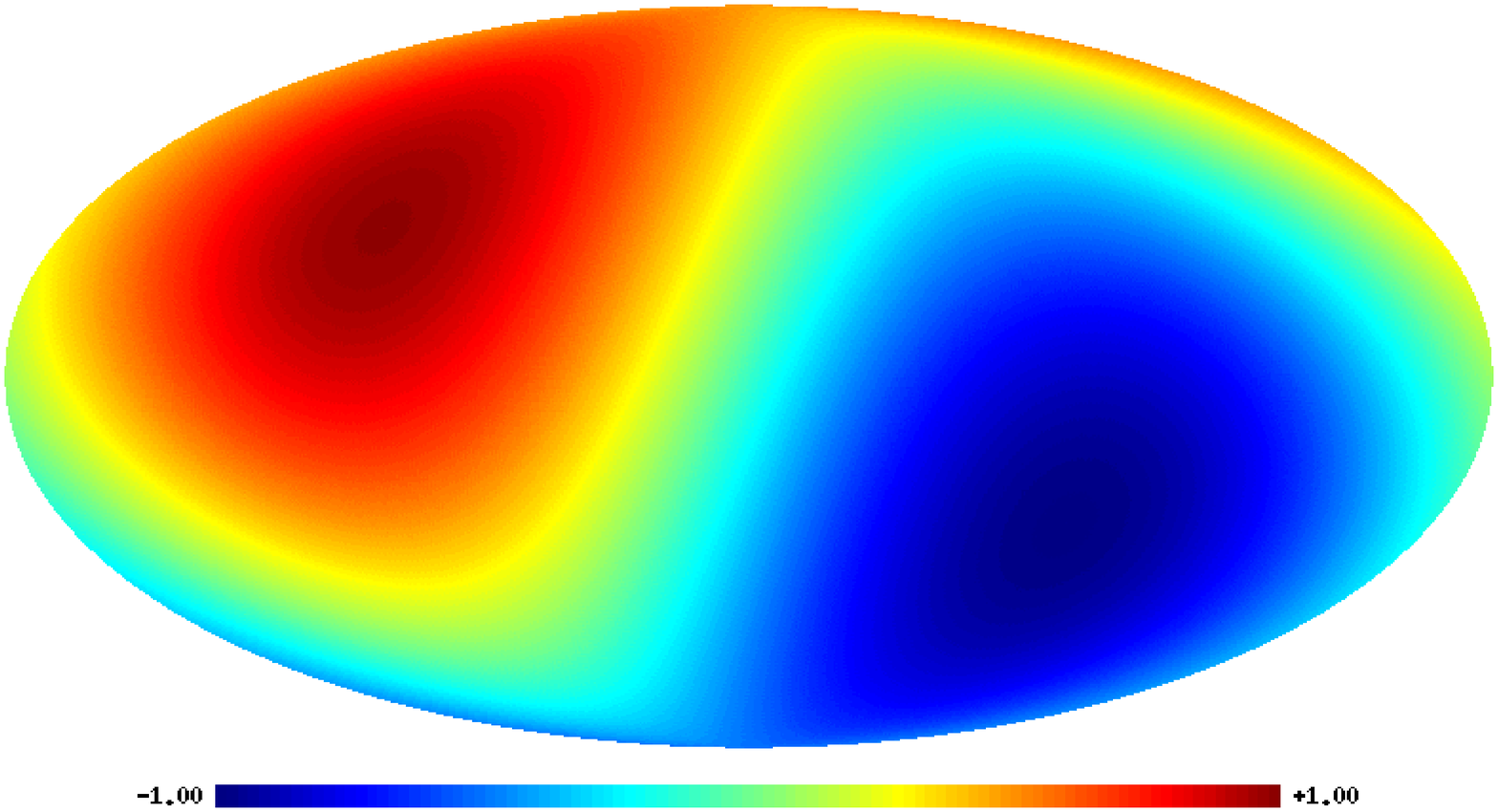}}
  \caption{Various explicitly power asymmetric templates used in our analysis. These are
          generated by modifying some of the \texttt{HEALPix} routines. The middle and the bottom
          ones are used in our studies to relate North-South ecliptic power asymmetry found by
          Eriksen \emph{et al.} (2004a) and the parity asymmetry that we are studying here.}
  \label{AsymTmplts}
\end{figure}

The even-odd multipole power asymmetry we are studying is a point inversion 
(PI) symmetry violation. So, we constructed some templates with explicit PI symmetry breakdown which may induce
a power excess in odd multipoles and incorporate them in generating our simulated raw maps.
In (Dobler \& Finkbeiner 2008a), an anomalous haze component was found in the WMAP data.
It could be that this anomalous haze has such asymmetry.
Also, there are sub-dominant foregrounds in microwave frequency region which are not well characterized yet
(Kogut et al., 1996; de Oliveira-Costa et al., 2002; Dobler \& Finkbeiner 2008b).
However, instead of making any such identifications here, we pursue the analysis including these new
templates as some hitherto unknown components. A readily conceivable pattern with such a point symmetry violation
is a hemispherical power asymmetry.  This template is shown in Fig. [\ref{AsymTmplts}] at the top.
We scale this template by a small factor before
including them in simulated raw data production pipeline so that it's contribution stays
sub-dominant and that the simulated raw maps conform visually with the accumulated raw data from
observations. The results presented in this section were obtained using 350 simulations. Since, we were
expecting to find a lower significance of parity asymmetry in the presence of asymmetric
foregrounds, 350 simulations are sufficient to probe it up to $3\sigma$ CL.

\begin{figure}[h]
 \centering
 \includegraphics[angle=-90,width=0.84\textwidth]{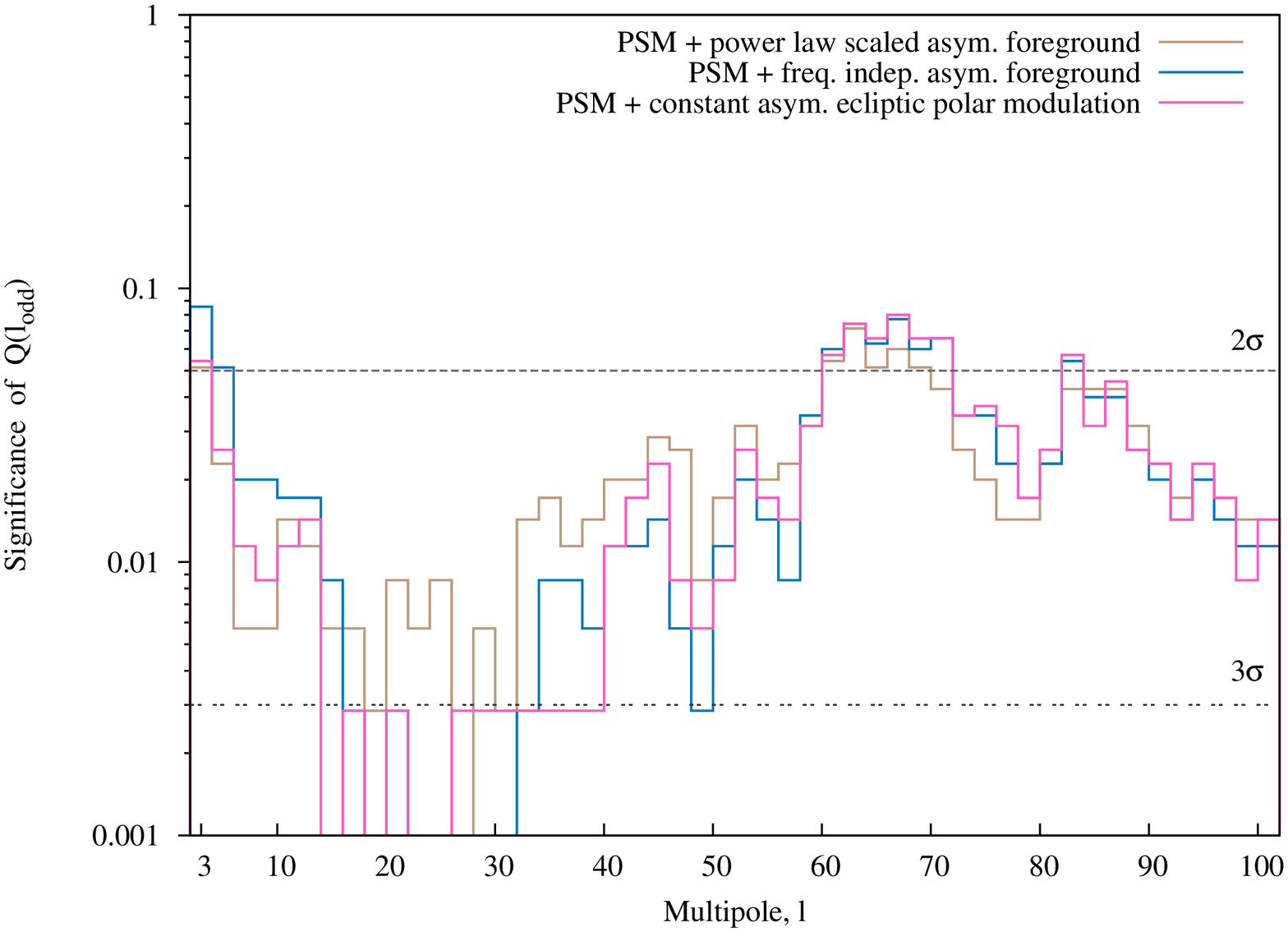}
 \caption{The $p-$value plot from the simualted ensemble set generated including our hitherto unknown
          foreground maps with explicit point inversion asymmetry using the $Q(l_{odd})$ statistic.
          The significances given here are for (1) power law scaled PI asymmetric map, (2) a constant
          power asymmetric map and (3) a foreground component which has explicit power asymmetry
          in its even and odd multipoles.}
 \label{Signf_OSasymFG}
\end{figure}

We used this template in two ways, one in which this template is scaled from K-band to
W-band of WMAP by a frequency dependent power law function and in the other instance as a
constant asymmetric foreground component, constant in all frequency channels.
In the former case, we chose the scaling factor to be $1/15^{th}$ the monopole intensity of synchrotron
from PSM at $23$GHz (99$\mu K$). It is further scaled to W-band following rigid frequency scaling
(Bouchet \& Gispert, 1999) with a steep spectral index as $F(\nu) = F(\nu_0)(\nu_0/\nu)^{2.8}$, where $F(\nu_0)$ is the
intensity distribution of a foreground component at a reference frequency $\nu_0$ extrapolated to another
frequency $\nu$. We chose a large spectral index so that this effect dies of at higher frequencies (V or W bands)
where CMB is supposed to be less contaminated by the foregrounds.
These templates, generated at five frequency bands of WMAP, 
were added to raw maps and convolved with
appropriate beam function. In the latter case, where this asymmetric map is added as a fixed power in each pixel
across all bands,
it is scaled by $1/30^{th}$ the synchrotron monopole intensity at $23$GHz. The scaling factor is chosen such that
this excess power will stay lower than the three dominant foregrounds in each channel.
The results are presented in Fig. [\ref{Signf_OSasymFG}] using our statistic.

We find that the significance decreases, but only a little in both the cases. These probability
estimates are similar to either pure maps or cleaned maps with only the three dominant foreground components
from PSM. We also used exponentially scaled prefactors for this template and found no increase
in probability, as a test for foreground components which may not be following polynomial scaling laws.

Then, we generated another template which is explicitly asymmetric in power between even and odd multipoles.
We generated $a_{lm}$'s with non-zero values for only $a_{l0}$'s and zero otherwise
and that the asymmetry dies (exponentially) with increasing $l$. The generated map is also shown
in Fig. [\ref{AsymTmplts}], in the middle. Eriksen \emph{et al.} (2004a) found a hemispherical
power asymmetry between the north and south ecliptic hemispheres. Motivated by the similarity of the
that North-South power asymmetry and PI symmetry violation, we explore whether there is any
relation between these two phenomena. So, we rotate this explicitly even-odd power asymmetric map
into ecliptic poles. This template is scaled by $1/20^{th}$ the monopole intensity of synchrotron
map from PSM and added it as a constant ecliptic dipolar power excess. Here we used a slightly higher
power to scale this template compared to the earlier constant hemispherical asymmetric template.

\begin{figure}[h]
 \centering
 \includegraphics[angle=-90,width=0.84\textwidth]{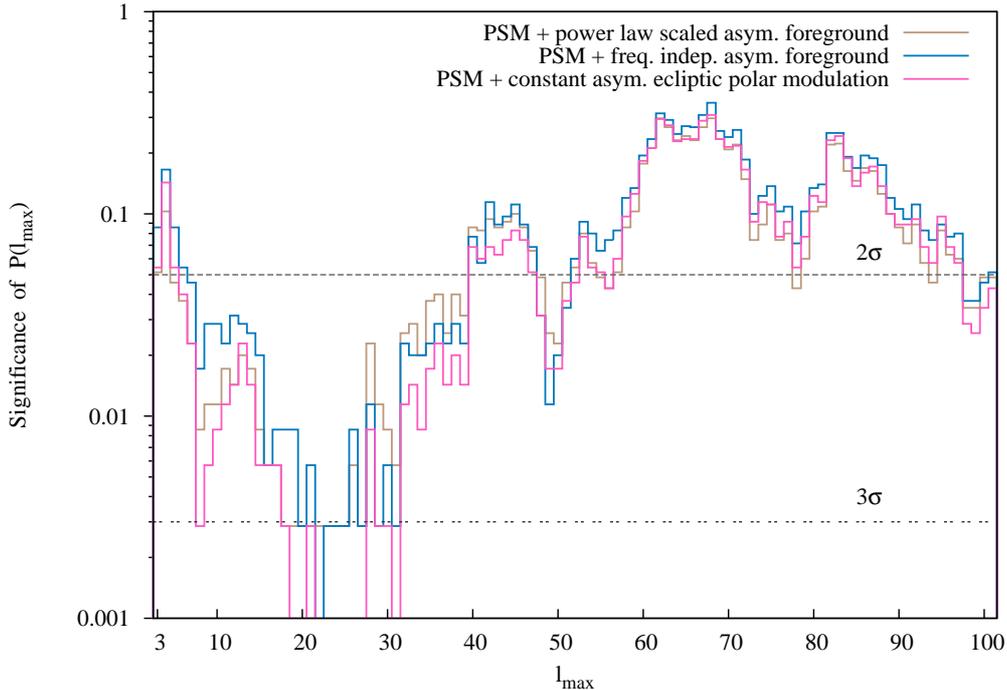}
 \caption{Same as Fig. [\ref{Signf_OSasymFG}], but for the $P(l_{max})$ statistic}
 \label{Signf_KNasymFG}
\end{figure}

Again we find no change in significance. It only decreases marginally. To find any significant effect
with this template, we had to add it at unrealistically high levels. 
 With low
intensity level, we do not find any increase in probability. The significance estimates with
our statistic, $Q(l)$, for this dipolar modulation are also shown in Fig. [\ref{Signf_OSasymFG}].
In Fig. [\ref{Signf_KNasymFG}], we presented the $p-$value estimates using the $P(l_{max})$ statistic.

\subsection{Multiplicative modulation}

So far we explored the possibility of only ``additive'' modulations to the data.
Now, we also consider ``multiplicative'' modulations
(Gordon et al., 2005; Eriksen et al., 2007b; Bunn \& Bourdon, 2008; Hanson \& Lewis, 2009)
which can break PI symmetry. The proposed modulation to the CMB is
\begin{equation}
 \Delta T(\hat{n}) = \Theta(\hat{n})(1+A\hat{\lambda} \cdot \hat{n})\,,
 \label{eriksenDipole}
\end{equation}
where $A$ is the modulation amplitude and $\hat{\lambda}$ is the preferred direction.
We chose $\hat{\lambda}$ to lie along the axis of ecliptic poles. Thus, we generated
a dipole modulation map of Eq. [\ref{eriksenDipole}] also shown in Fig. [\ref{AsymTmplts}],
at the bottom.

\begin{figure}[h]
  \centering
  \includegraphics[angle=-90,width=0.84\textwidth]{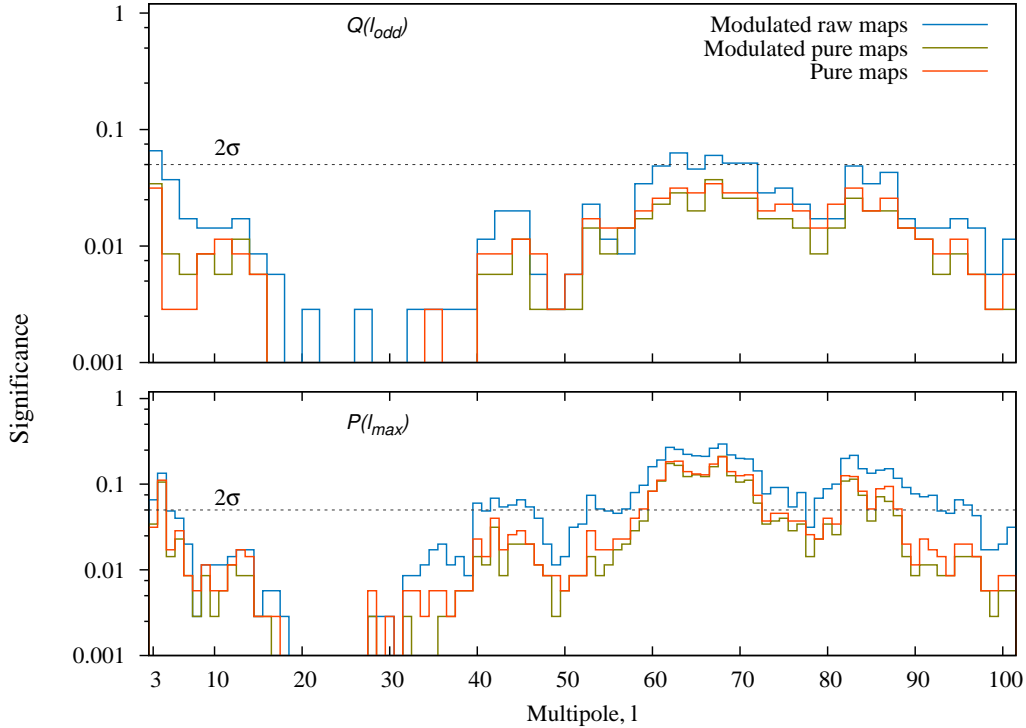}
  \caption{The $p-$values obtained after applying a dipolar multiplicative 
modulation to pure maps and raw maps before cleaning
           is shown here. The parity statistics of modulated pure maps are not different from the pure maps themselves.
           But the modulated raw maps show a slight decrease in the significance
of power asymmetry. 
           }
 \label{Signf_KN_OS_EclipticDipoleModul}
\end{figure}

Even with an amplitude of $A=0.3$ we find that it cannot induce a particular parity preference.
The results from modulated pure maps are shown in Fig. [\ref{Signf_KN_OS_EclipticDipoleModul}]. Also shown there are the
pure maps estimates for comparison.
When the same modulation is applied on the raw maps, there is an enhancement in
the parity asymmetry in the cleaned map, Fig. [\ref{Signf_KN_OS_EclipticDipoleModul}]. This enhancement of power asymmetry in modulated raw
maps suggest a measurement artifact rather than any thing fundamental to CMB radiation. We point out that
we are using $A=0.3$ which is relatively large amplitude for modulation. 
Even with such a large amplitude, 
we don't find much change in the statistic's values of modulated pure maps compared to pure maps themselves.

\section{IPSE and others}
\label{section5}

\begin{figure}[h]
 \centering
 \includegraphics[angle=-90,width=0.84\textwidth]{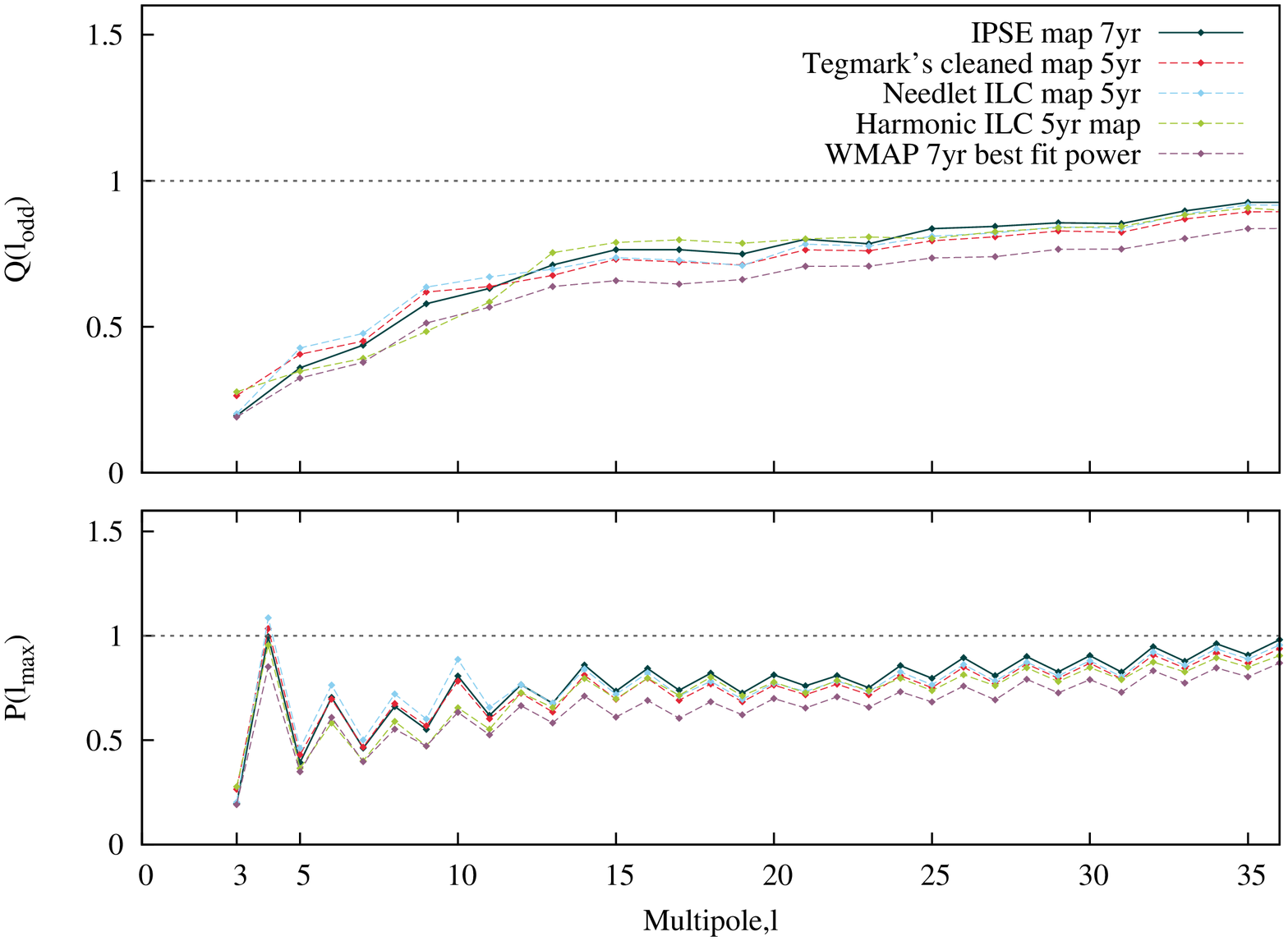}
 \caption{The parity statistic values using $Q(l)$ (top panel) and 
$P(l)$ (bottom panel) applied to cleaned maps, using several different procedures,
are shown here. The maps used are (1) IPSE map, (2) Tegmark's five year cleaned map,
 obtained using the procedure described in Tegmark, de Oliveira-Costa \& Hamilton, 2003, (3) Needlet ILC five year map and (4) Harmonic ILC five year map. Also plotted are
          the power asymmetry statistic values for WMAP's seven year best fit temperature power
          spectrum for comparison. All these maps show similar levels of parity asymmetry,
          but do not agree with WMAP's best fit data.}
\label{OS_KN_ipse}
\end{figure}

\begin{figure}[h]
 \centering
 \includegraphics[angle=-90,width=0.84\textwidth]{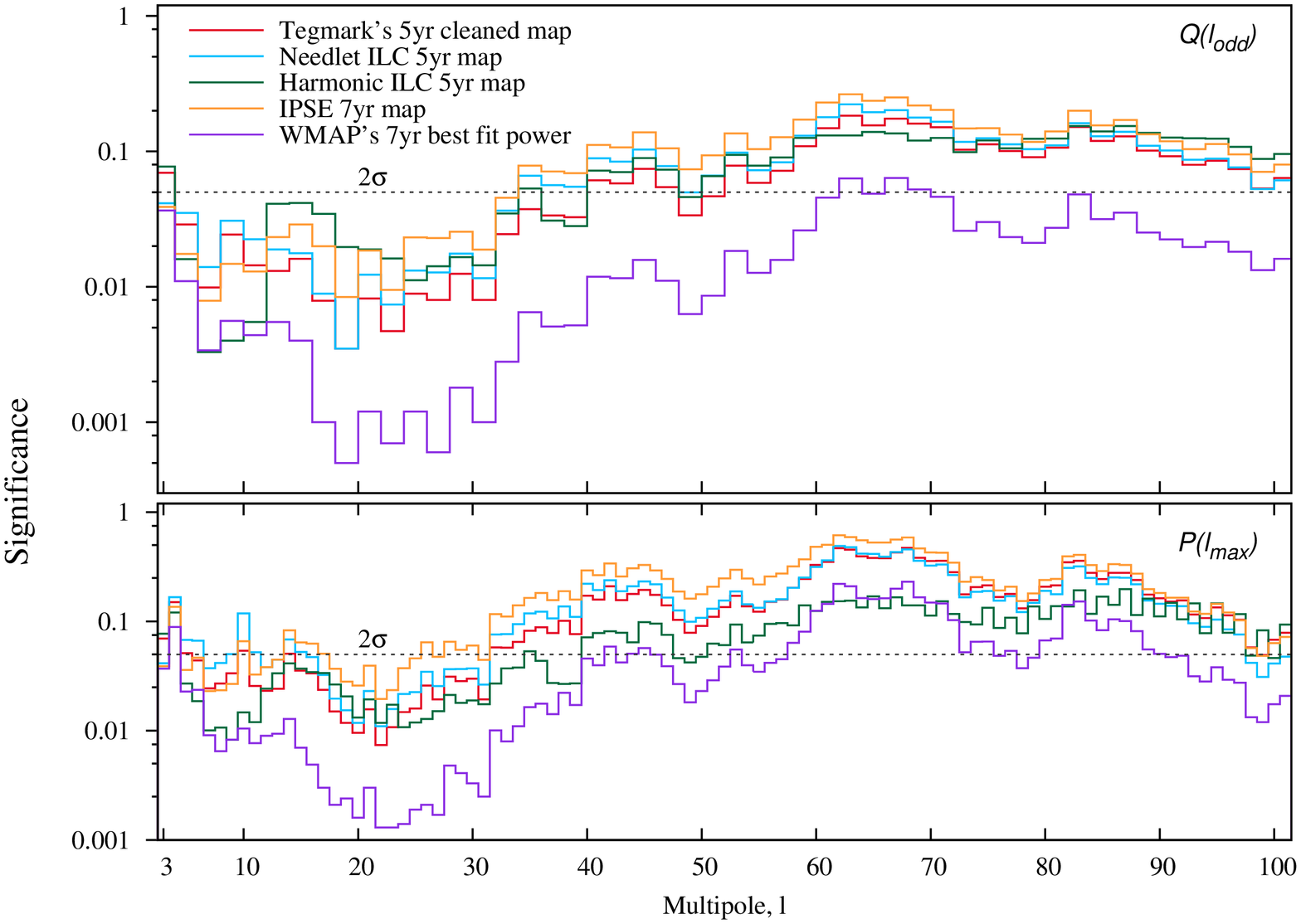}
 \caption{The $p-$values for parity asymmetry for the cleaned maps 
using several different procedures 
(see text). 
          }
 \label{Signf_OS_KN_ipse}
\end{figure}

We next test the signal of parity asymmetry in maps cleaned by
IPSE and several other procedures. In the case of IPSE we perform a
full sky cleaning of the temperature raw data from WMAP's seven year data release.
The power spectrum is computed from full sky cleaned map up to $l=10$ and used
pseudo$-C_l$ estimator at higher $l$ after applying
WMAP's $KQ85yr7$ mask. Later we estimate the power at low$-l$ also from 
masked sky. This allows us to determine the how the parity statistic
is influenced by masking. The $Q(l)$ and $P(l_{max})$
values in the range $l=[2,101]$ for this power spectrum are shown in Fig. [\ref{OS_KN_ipse}]
and the probability estimates are shown in Fig. [\ref{Signf_OS_KN_ipse}]. These $p-$values
are computed from $10,000$ simulated pure maps.

We see that the IPSE cleaned data does not show significant power asymmetry. It was surprising
to find this result, given that the WMAP seven year best fit power spectrum is found to be
highly anomalous. So, we also applied the two statistics to other
cleaned maps available to us. The maps
considered here are (1) cleaned map from WMAP's five year raw
data\footnote[6]{http://space.mit.edu/home/tegmark/wmap/ or https://www.cfa.harvard.edu/$\sim$adeolive/gsm/index.html} obtained by using the procedure given
in 
 Tegmark, de Oliveira-Costa \& Hamilton, 2003,
(2) Needlet ILC map\footnote[7]{http://www.apc.univ-paris7.fr/APC\_CS/Recherche/Adamis/cmb\_wmap-en.php}
of (Delabrouille \emph{et al.}, 2009) using WMAP's five year data and
(3) Harmonic ILC\footnote[8]{http://www.nbi.dk/$\sim$jkim/hilc/} of (Kim, Naselsky \& Christensen, 2008),
which is also produced from WMAP five year data. We note that the maps (1) and (2) are available at
resolution of W-band, just like IPSE cleaned map. But the Harmonic ILC map is available at $1^o$ resolution. In all these cases the power is obtained from 
the full sky cleaned map up to $l=10$ and pseudo$-C_l$ estimator for 
$l>10$, as in the case of IPSE.
The parity asymmetry statistic values of these maps are also shown in Fig. [\ref{OS_KN_ipse}].
As can be seen, all these maps give results close to each other, but do
not agree with those obtained using the WMAP seven year
best fit temperature power spectrum. Hence their significances
are similar to IPSE cleaned data, as shown in Fig. [\ref{Signf_OS_KN_ipse}].

\begin{figure}[h]
 \centering
 \includegraphics[angle=-90,width=0.84\textwidth]{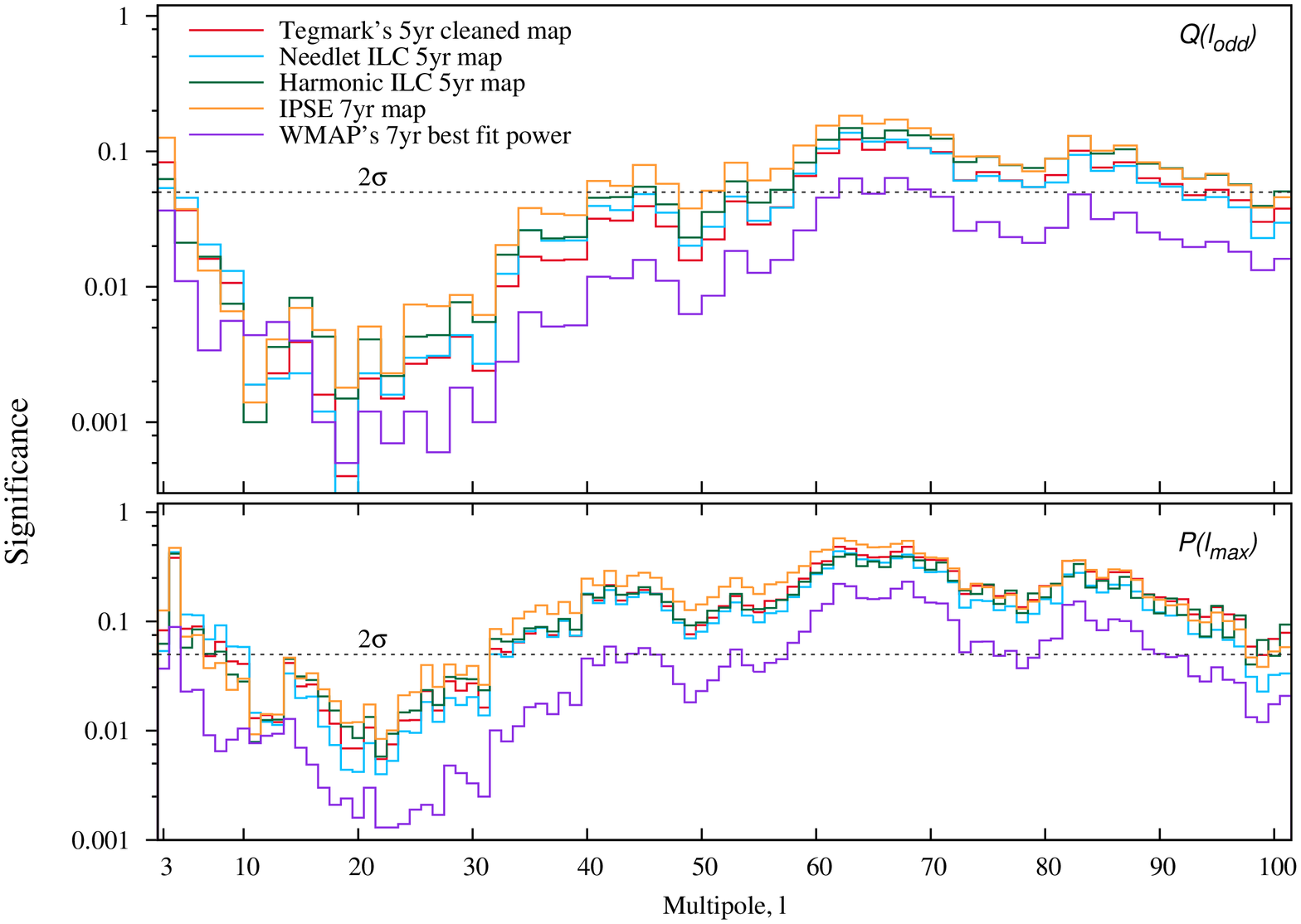}
 \caption{The significance using the pseudo-$C_l$ estimator. The
          different maps mentioned in the legend are masked using $KQ85yr7$ mask.}
 \label{Sign_OS_KN_ipse_cut}
\end{figure}

The power spectrum in all the cases analysed in this section is obtained 
from full sky up to $l=10$ and masked sky for $l>10$. This is in contrast
to the WMAP best fit power spectrum which uses masked sky over the entire 
multipole range. Hence it is useful to determine how the results for
the maps considered in this section change if we use a pseudo$-C_l$ estimator
for $l\le 10$ also.
In order to estimate parity asymmetry using masked sky over the entire multipole
range, we applied the WMAP's $KQ85yr7$ mask on all these cleaned
maps including our IPSE map. We then obtained the corresponding
full sky pseudo-$C_l$ values for these maps.
The corresponding significance of the parity asymmetry, using both the
statistics, is shown in Fig. [\ref{Sign_OS_KN_ipse_cut}].
With our statistic, we find that all the different maps are relatively close
to the WMAP's best fit power spectrum.
The $P(l_{max})$ statistic also shows a higher significance in comparison 
to the full sky power spectra.
Hence, we find that, pseudo$-C_l$ estimator, recovered from masked clean 
maps, reveals the presence of
anomalous parity asymmetry in these maps.
This is most likely due to the fact that the
full sky has many heavily contaminated regions where the cleaning may not
be very efficient. By masking such regions we hope to get a better estimate 
of the true power spectrum of the CMB signal.
It has been found earlier that some of the large angle anomalies
disappear in cut sky maps
(Bielewicz et al., 2005; Bernui et al., 2007; Efstathiou, Ma \& Hanson, 2010;
Pontzen \& Peiris, 2010; Copi et al., 2011).
It is therefore encouraging that in the present case the signal is 
enhanced rather than diminished when we use masked sky.

\begin{figure}[h]
\centering
\includegraphics[angle=-90,width=0.84\textwidth]{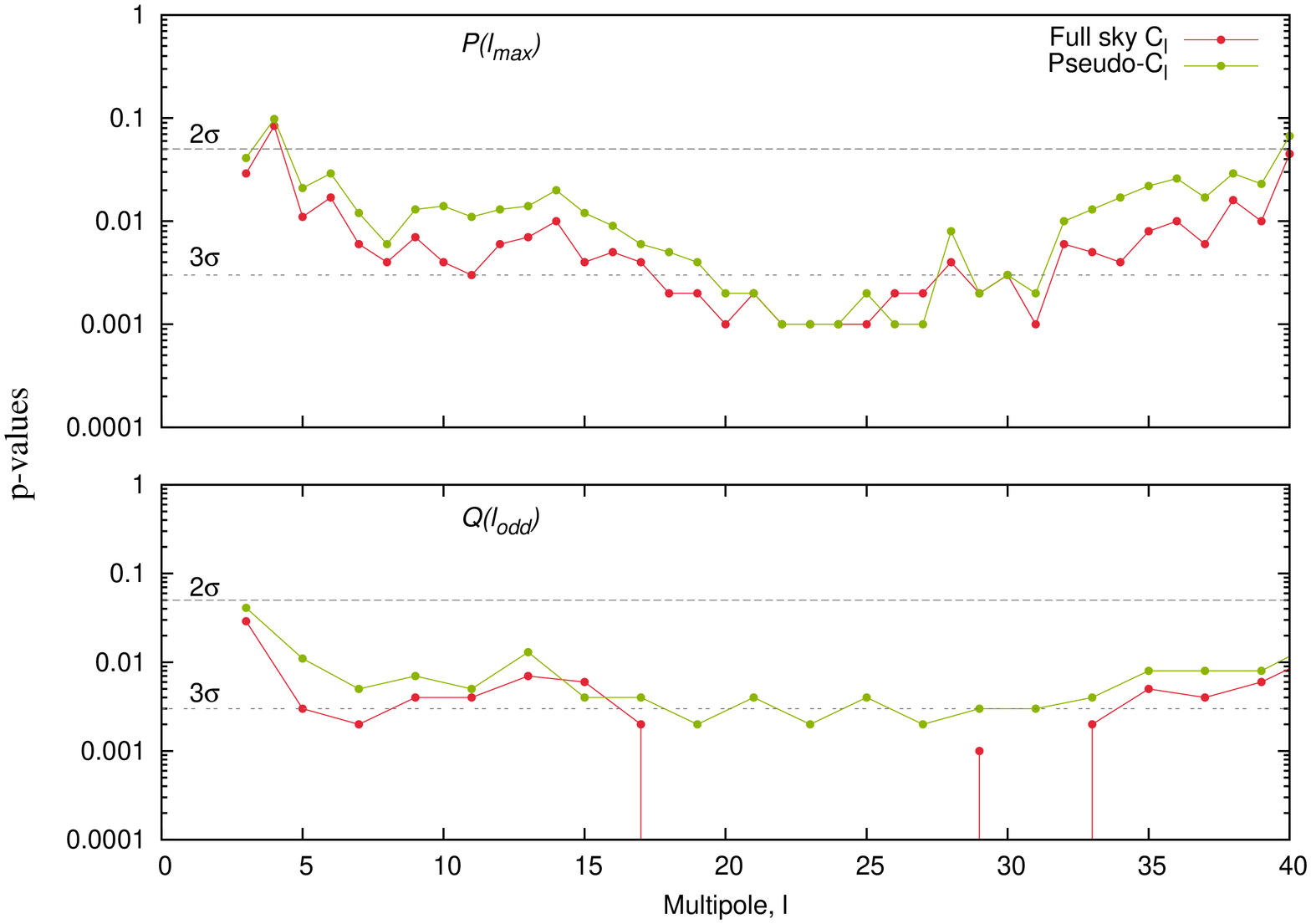}
\caption{The significance for the WMAP best fit power spectrum 
using masked sky random realizations of pure CMB. The results where the
power spectrum of random realizations is estimated from full sky maps is
shown for comparison.}
 \label{Signf_OS_KN_fs_pseudo}
\end{figure}

The fact that a pseudo$-C_l$ estimator gives a higher significance of parity
asymmetry in comparison to the power spectrum obtained from full sky may lead
one to suspect that the process of masking itself might generate some
signal of parity asymmetry. In order to study this possibility we determine
the significance of parity asymmetry in WMAP best fit power spectrum by
using simulated masked random realizations of pure CMB. We use the
$KQ85yr7$ for this purpose.  
The resulting 
significance levels for both the statistics are shown in Fig. [\ref{Signf_OS_KN_fs_pseudo}]. For
comparison we also show the results for the case when the power spectrum
of the simulated maps are obtained from full sky.  
We find that if we use the masked sky pseudo$-C_l$ estimator for the random
samples, the significance level for parity asymmetry is slightly lower for
both the statistics. 
Though there is a net rise in $p-$values due to masking, the
relative change is marginal/low and the
signal of anomalous parity asymmetry is still present.

\subsection{Statistical significance using the ILC procedure for foreground
removal}

So far we have used the WMAP best fit power spectrum in our analysis.
We computed the statistical significance by comparing the statistic for the 
best fit power with that obtained from the randomly generated pure CMB
maps as well as simulated, foreground cleaned CMB maps. 
The simulated CMBR maps were cleaned using the IPSE procedure.  
We have also studied the parity asymmetry in other cleaned maps obtained using
various procedures.
In this section we compute the statistical significance 
using the ILC procedure for foreground removal. The ILC procedure is used
both for estimating the statistic for the WMAP data as well as for cleaning
the simulated maps.

\begin{figure}[h]
 \centering
 \includegraphics[angle=-90,width=0.84\textwidth]{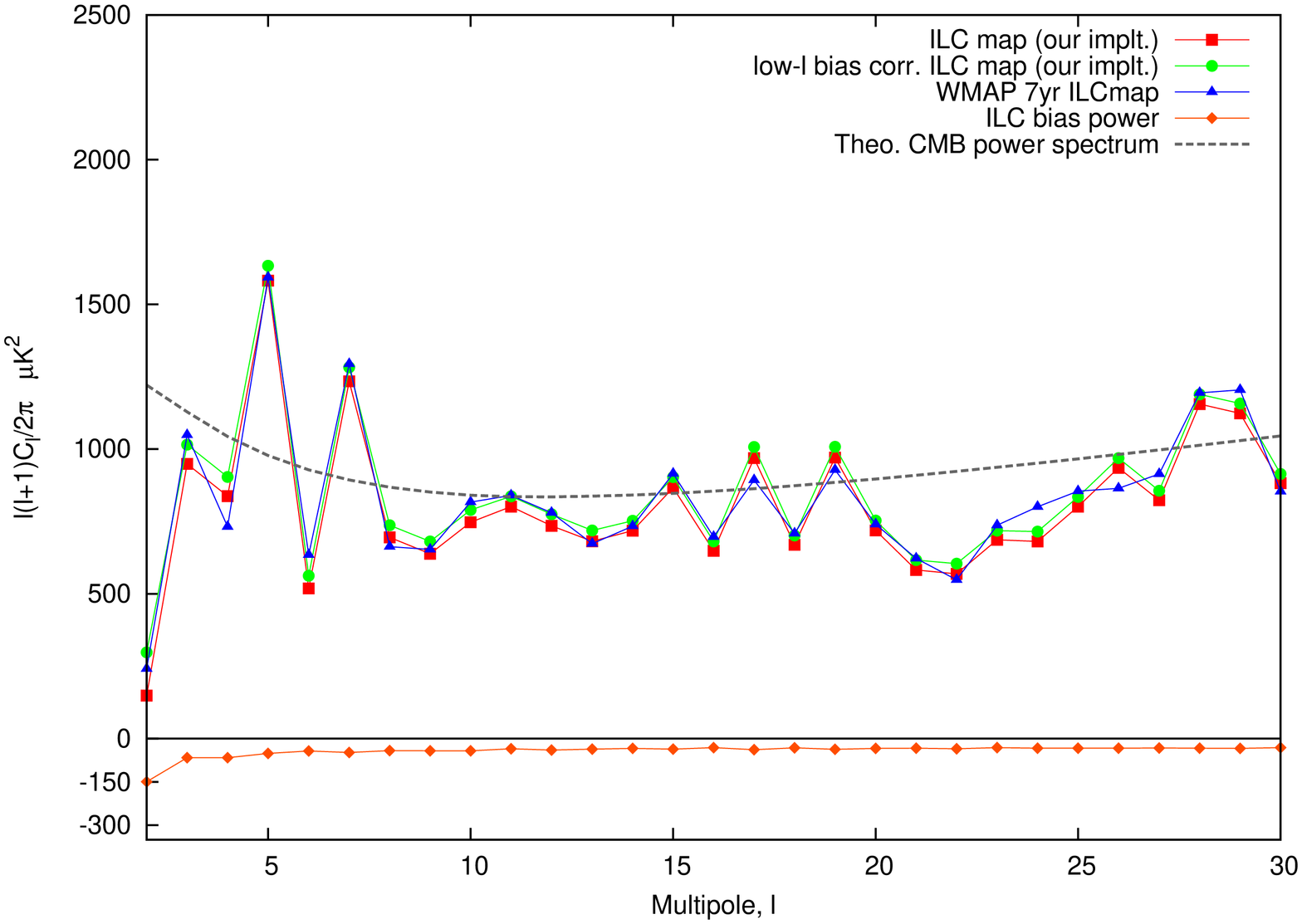}
 \caption{Power spectrum of ILC cleaned map from our implementation and from the WMAP seven year ILC map is plotted here.
          The curve below zero of y-axis is the bias in the power spectrum from ILC cleaning
          method. This is computed as an average over $600$ simulated pure maps and clean maps generated at $1^o$ resolution. Also
          shown is the best fit theoretical CMB temperature power spectrum for comparison. Our implementation
          and WMAP cleaned ILC map agree with each other. Any difference could be due to our bias correction
          map estimated using PSM. WMAP uses MEM foreground templates for generating the
          foreground bias map.}
 \label{ILCpower}
\end{figure}

In the ILC procedure, foregrounds are removed by making linear combination
of maps at different frequencies in pixel space. Maps at different frequencies are
smoothed to a common resolution of $1^o$ and added with suitable weights
to minimize the foreground power in the combined map.  
The details of the ILC procedure are given in
(Bennett et al., 2003c).
In Fig. [\ref{ILCpower}] we compare the power extracted by 
our implementation of the ILC 
procedure with that obtained by WMAP. We find that the two are in 
good agreement with one another. 

In earlier papers, it has been shown that both the IPSE and ILC procedures
are expected to have some bias at low $l$. The foreground cleaning procedure
removes some extra power and hence the extracted signal is lower in comparison
to the real signal. This effect is dominant at low multipoles $l=2,3$. 
Here we compute this bias for ILC using 600 simulations.  
The extracted bias is also shown
in Fig. [\ref{ILCpower}]. 
As expected we find a negative bias at low$-l$ in power spectrum estimation 
(Hinshaw et al., 2007; Saha et al., 2008; Chiang, Naselsky \& Coles, 2009). However we find that the bias
is much smaller in comparison to that obtained using IPSE.
The final power spectrum after removing this negative bias is also shown
in Fig. [\ref{ILCpower}].

\begin{figure}[h]
 \centering
 \includegraphics[angle=-90,width=0.84\textwidth]{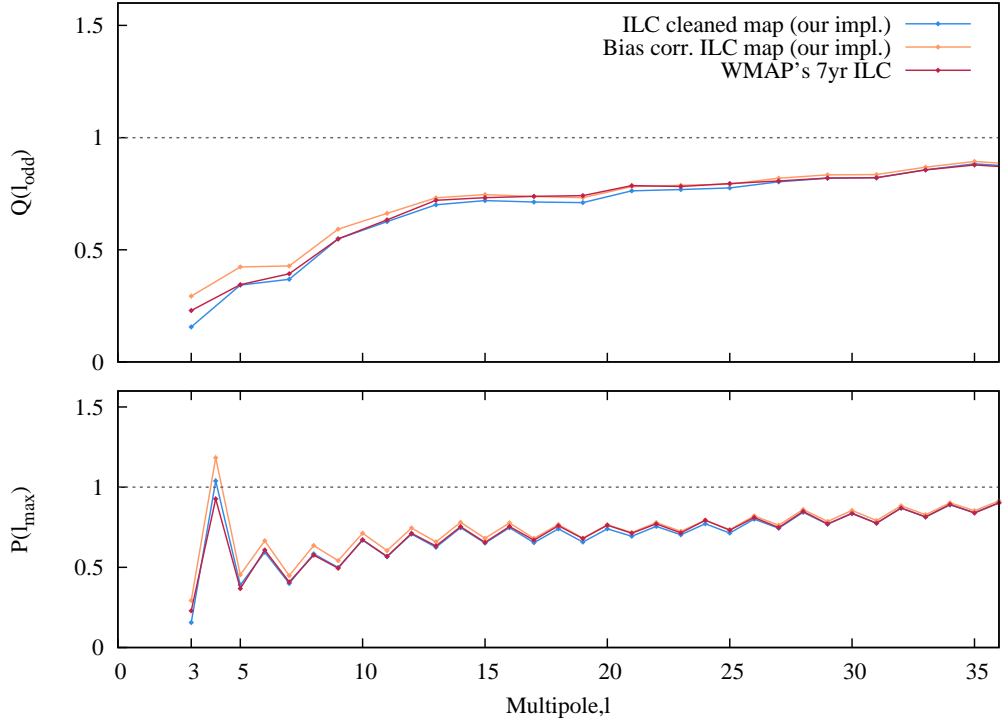}
 \caption{The two statistics applied to ILC cleaned WMAP seven year data. The statistics are shown
          for the ILC cleaned map as cleaned by us, bias corrected ILC power and WMAP's
          ILC 7yr map.}
 \label{ilc_twostats}
\end{figure}

\begin{figure}[h]
 \centering
 \includegraphics[angle=-90,width=0.84\textwidth]{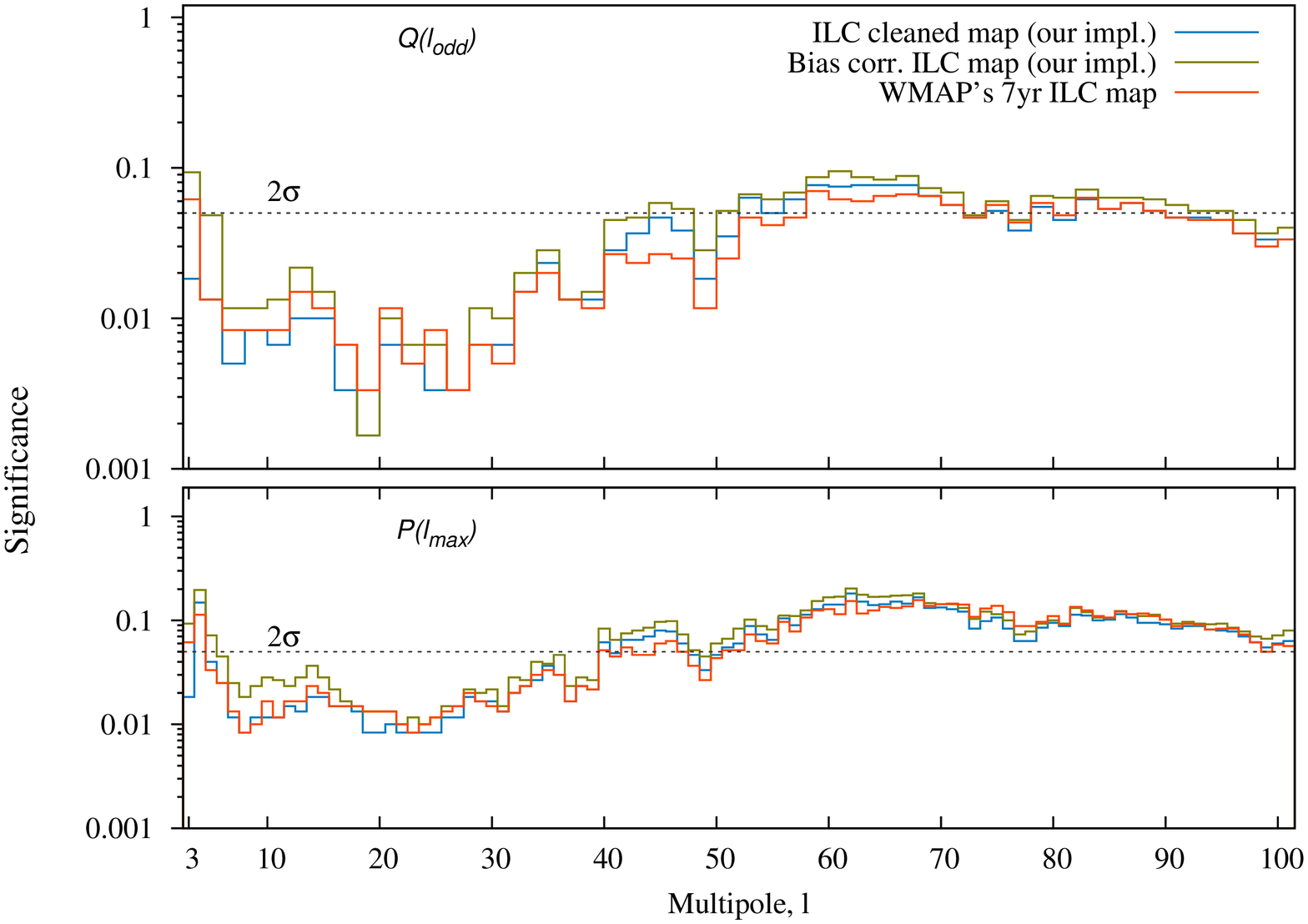}
 \caption{The $p-$values of the power asymmetry from the ILC cleaned data.
 Shown here
          are the probability estimates for the two statistics from ILC map as cleaned
          by us, bias corrected ILC map from our implementation and the WMAP team
          cleaned seven year ILC map. As seen, the probability of the parity asymmetry
          is lower compared to the WMAP's seven year best fit power. However
 the probability rises slightly  
         after correcting for the low$-l$ bias. 
          }
 \label{signfILC}
\end{figure}

In Fig. [\ref{ilc_twostats}] we have shown both the statistics computed for the
ILC cleaned map. The corresponding statistical significance using the
two statistics is shown in Fig. [\ref{signfILC}].
The results were presented for the WMAP seven year ILC map, the ILC map obtained
by us as well as the low$-l$ bias corrected ILC power. We find that the
statistical significance of all the three maps are comparable to one another.
We note that the ILC map is reliable on angular scales greater than $10^o$ (Bennett et al., 2003c).
It is available at a resolution of $1^o$ and so is HILC map of (Kim, Naselsky \& Christensen, 2008). So, it will not be
meaningful to assess the parity preference in that data at high $l$, even if it shows such an
asymmetry, where it's power spectrum deviates away from the theoretical CMB power spectrum.

\section{Parity significance with low$-l$ cuts}
\label{section6}

From Fig. [\ref{KN_OS_bestfit}], we see that the values of both the statisics,
$Q(l)$ and $P(l_{max})$, are much lower at small $'l\,'$.
This is also reflected in Fig. [\ref{Signf_KNstat}] and Fig. [\ref{Signf_ourstat}]
where the $p-$values are found to be relatively insignificant at
low$-l$. Also, as mentioned earlier, many studies found that the low$-l$
are associated with various anomalies. Hence it is reasonable 
to assess the parity asymmetry
neglecting some of the low$-l$ multipoles. 
In order to get a better insight into this parity asymmetry issue and avoid any
``anomalous low$-l$'' concerns, we discard some low$-l$ values in this analysis and compute
both the statistics for WMAP data and assess it's significance. Thus our 
statistic now becomes,
\begin{equation}
 Q(l_{odd}) = \frac{2}{l_{odd} -l_{cut} + 1}\sum_{l=l_{cut}}^{l_{odd}} \frac{\mathbb{C}_{l-1}}{\mathbb{C}_l}\,,
\end{equation}
where $l_{cut}$ is any odd $l>3$ and the summation is again over all odd multipoles $\leq l_{odd}$.
We implement a similar $l-$cut for $P(l_{max})$ at low$-l$ in computing $P^+$ and $P^-$.

\begin{figure}[h]
 \centering
 \includegraphics[angle=-90,width=0.84\textwidth]{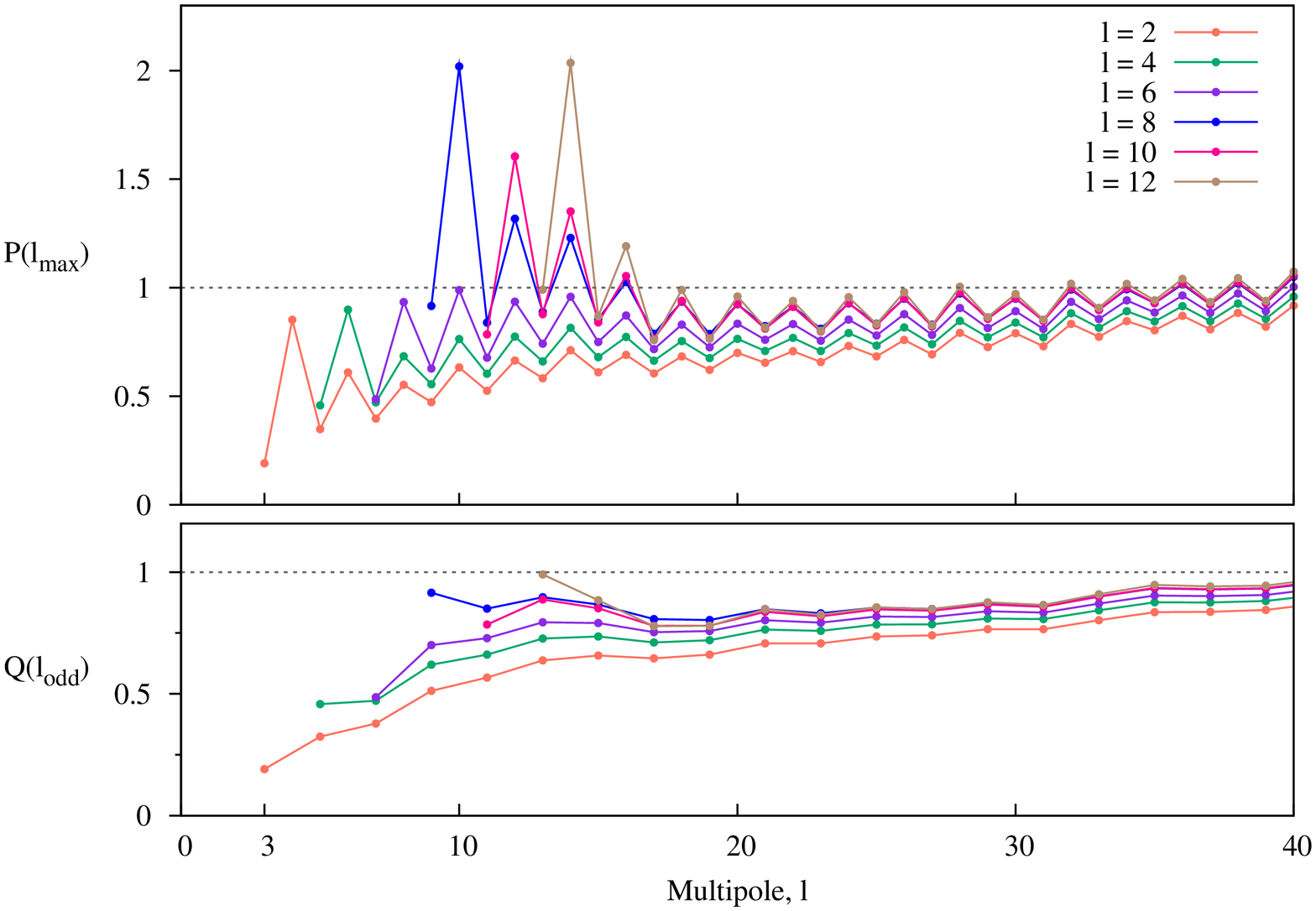}
 \caption{The parity preference estimators $Q(l)$ and $P(l_{max})$ with
          various low$-l$ cuts as applied to WMAP seven year best fit power spectrum.
          We see that the estimators steadily rise close to one with increasing low$-l$ cut.}
 \label{data7yr_lcuts}
\end{figure}

\begin{figure}[h]
 \centering
 \includegraphics[angle=-90,width=0.84\textwidth]{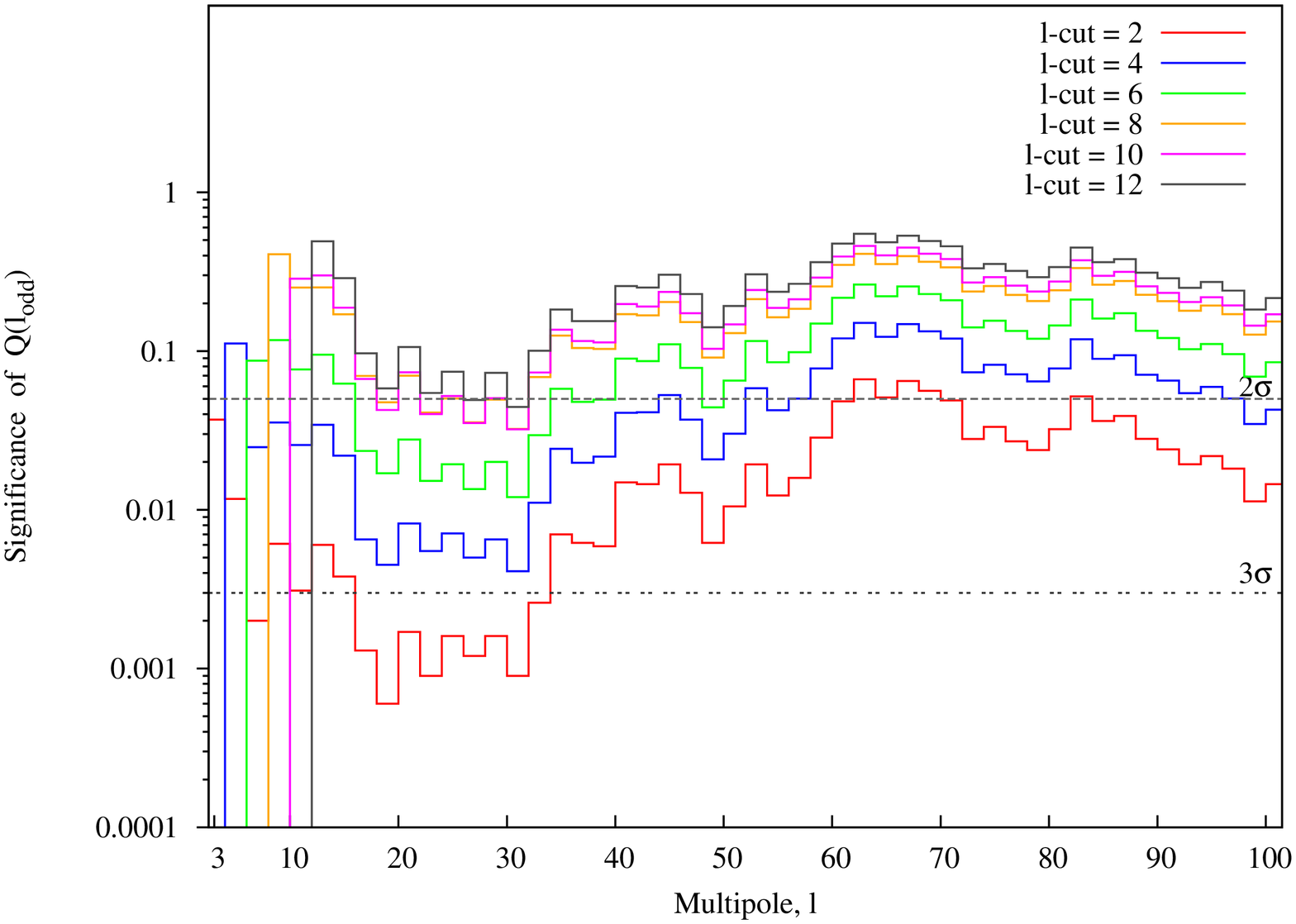}
 \caption{The $p-$ values of our parity preference estimator, $Q(l)$,
          as applied to WMAP's best fit power with various low$-l$ cuts, using an ensemble
          of 10,000 pure CMB realizations. The significance falls below $95\%$ CL just by ignoring the first 6 multipoles.}
 \label{OS_lcut}
\end{figure}

\begin{figure}[h]
 \centering
 \includegraphics[angle=-90,width=0.84\textwidth]{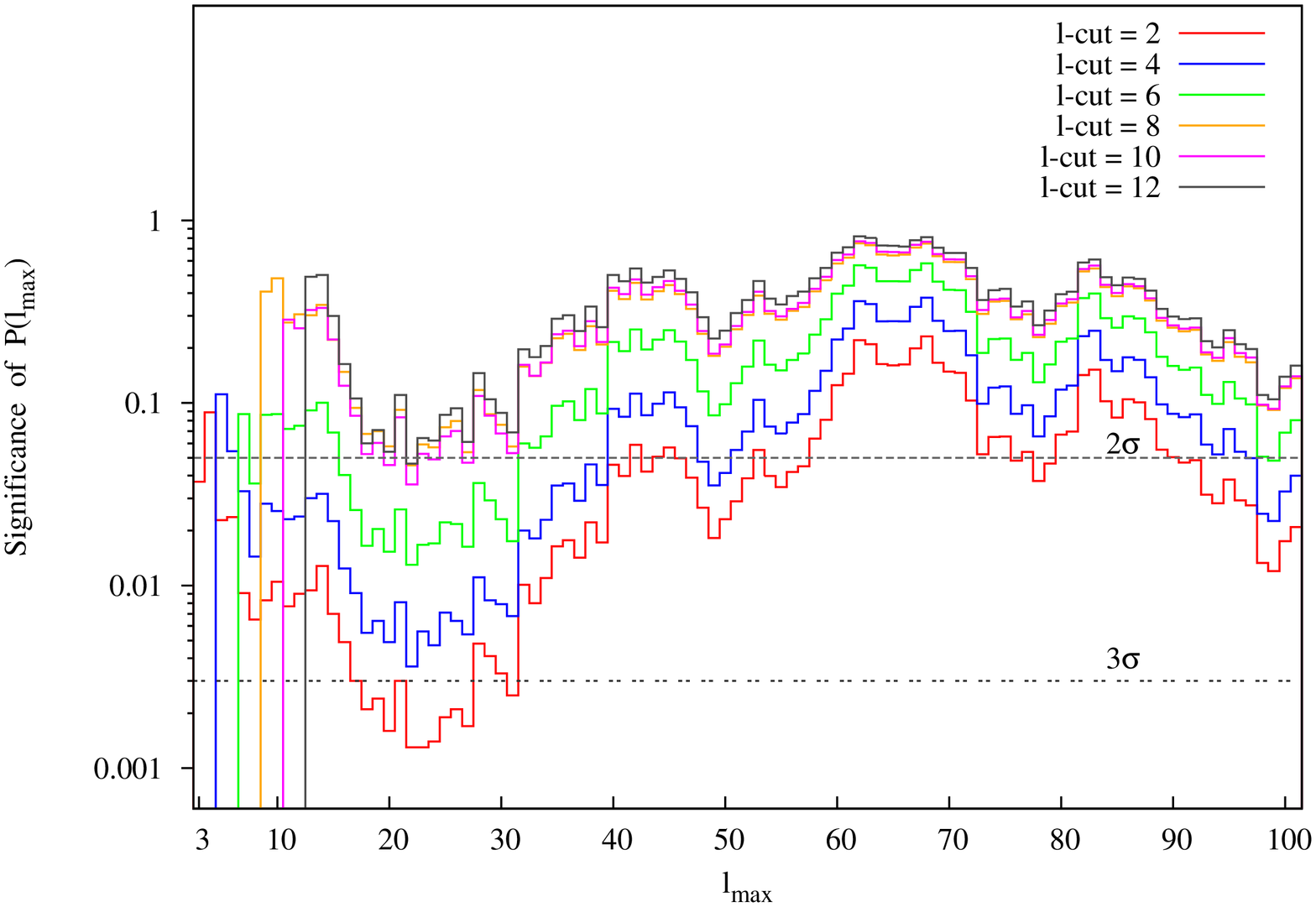}
 \caption{Same as Fig. [\ref{OS_lcut}] but for $P(l)$ statistic. Here too we find that
          the significance we found earlier disappears by ignoring the first 6 multipoles.}
 \label{KN_lcut}
\end{figure}

The result of applying the two statistics to WMAP best fit power with various low$-l$ cuts
is shown in Fig. [\ref{data7yr_lcuts}]. As can be seen, both the asymmetry statistics rises closer to one
with increasing multipole cuts.
With different low$-l$ cuts, the $p-$values at various $l$ in the range $l=[l_{cut},101]$ are computed
for both the statistics and the results are given in Fig. [\ref{OS_lcut}] and in Fig. [\ref{KN_lcut}].
We see from these figures that the significance of parity asymmetry immediately starts decreasing 
and falls below $2\sigma$ just by ignoring the first 6 multipoles ($l=2,..,7$). This
happens with both the statistics. It shows that the dominant contribution to the parity asymmetry
arises from a few low $l$ multipoles only. It raises the interesting possibility
that this might be related to the other low-$l$ anomalies seen in the
WMAP data. These might arise from a common physical origin.

\section{Conclusions}
\label{section7}

We have analysed the signature of 
  parity asymmetry, recently found in
WMAP's best fit temperature power spectrum, in considerable detail in the
multipole range $l=[2,101]$. For this purpose, we used the statistic,
$P(l_{max})$, introduced earlier as well as a new measure, $Q(l_{odd})$, 
which appears to be more sensitive. We confirm the signal of parity 
asymmetry at significance level of $3\sigma$. By comparing an ensemble of
simulated foreground cleaned maps with the WMAP best fit power spectrum,
we deduce that the observed parity asymmetry cannot be attributed to
foreground cleaning or residual foregrounds. Here we use the foreground
templates from PSM. The PSM may not
correctly model some sub-dominant or unknown foregrounds. Hence we created
some templates which explicitly violate parity and included them in our analysis
as both additive and multiplicative modulation to CMB. We again find that these cannot
explain the observed power asymmetry. The level of asymmetry present 
in data can be obtained by introducing an unrealistically large value
of these foreground components. Hence we find that we are unable to 
attribute the observed signal to foreground cleaning or residual foregrounds.  

We next tested the presence of this signal of parity asymmetry in several 
other foreground cleaned maps such as the IPSE map, cleaned map using 
the procedure of Tegmark, de Oliveira-Costa \& Hamilton, 2003, Needlet ILC map
and Harmonic ILC map. We find that these maps also show a signal of parity
asymmetry provided we use the pseudo$-C_l$ estimator after applying a mask 
in order to eliminate the heavily foreground contaminated regions. The   
significance level for these maps, however, is found to be not as high
as that in the case of WMAP best fit power.  
The ILC map also shows parity asymmetry with
 results closer to that obtained with the WMAP best fit 
power spectrum.

Finally, we tested the WMAP data for parity asymmetry by eliminating some of the
low$-l$ modes. The low$-l$ multipoles are known to show some anomalous
results such as low quadrupole power (Bennett et al. 2003b), 
alignment of various multipoles (de Oliveira-Costa et al., 2004) etc. 
Hence it is possible that the parity asymmetry might also get a large
contribution from these multipoles.
We found that the parity
asymmetry disappears by just ignoring the first six multipoles ($l=2,..7$).
Hence we conclude that the low-$l$ multipoles give dominant contribution
to the signal of parity asymmetry. It is, therefore, possible that 
all the low-$l$ anomalies,
including the parity asymmetry might have a common origin. 
\\
\\{\bf Acknowledgements} \\
We acknowledge the use of WMAP data available from NASA's LAMBDA site(http://lambda.gsfc.nasa.gov/).
We also used the publicly available \texttt{HEALPix} software (Gorski et al., 2005) for
map handling and also to extract relevant information from these maps.
We thank John P. Ralston for useful discussions. We thank Pavel Naselsky
for a useful communication. 

\bigskip

{\bf\large  References}

\begin{itemize}

\item[] A. Ben-David, E. D. Kovetz, N. Itzhaki, 2011, arXiv:1108.1702

\item[] C. L. Bennett et al., 2003a, ApJ, 583, 1

\item[] C. L. Bennett et al., 2003b, ApJS, 148, 1

\item[] C. L. Bennett et al., 2003c, ApJS, 148, 97

\item[] C. L. Bennett et al., 2011, ApJS, 192, 17

\item[] A. Bernui, B. Mota, M. J. Reboucas and R. Tavakol, 2007, Int. J. of Mod. Phys. D, 16, 411

\item[] P. Bielewicz, H. K. Eriksen, A. J. Banday, K. M. Gorski and P. B. Lilje, 2005, ApJ, 635, 750

\item[] F. R. Bouchet and R. Gispert, 1999, New Astron., 4, 443

\item[] E. F. Bunn and A. Bourdon, 2008, Phys. Rev. D, 78, 123509

\item[] L-Y. Chiang, P. D. Naselsky and P. Coles, 2009, ApJ, 694, 339

\item[] C. J. Copi, D. Huterer and G. D. Starkman, 2004, Phys. Rev. D, 70, 043515

\item[] C. J. Copi, D. Huterer, D. J. Schwarz and G. D. Starkman, 2007, Phys. Rev. D, 75, 023507

\item[] C. J. Copi, D. Huterer, D. J. Schwarz and G. D. Starkman, 2010, Advances in Astronomy, 847541

\item[] C. J. Copi, D. Huterer,  D. J. Schwarz and G. D. Starkman, 2011, arXiv:1103.3505

\item[] A. de Oliveira-Costa et al., 2002, ApJ, 567, 363

\item[] A. de Oliveira-Costa, M. Tegmark, M. Zaldarriaga and A. Hamilton, 2004, Phys. Rev. D, 69, 063516

\item[] A. de Oliveira-Costa and M. Tegmark, 2006, Phys. Rev. D, 74, 023005

\item[] J. Delabrouille et al., 2009, A\&A, 493, 835

\item[] G. Dobler and D. P. Finkbeiner, 2008a, ApJ, 680, 1222

\item[] G. Dobler and D. P. Finkbeiner, 2008b, ApJ, 680, 1235

\item[] G. Efstathiou, 2003, MNRAS , 346, 2, L26

\item[] G. Efstathiou, Y-Z. Ma and D. Hanson, 2010, MNRAS, 407, 2530

\item[] H. K. Eriksen, F. K. Hansen, A. J. Banday, K. M. Gorski and P. B. Lilje, 2004a, ApJ, 605, 14

\item[] H. K. Eriksen, D. I. Novikov, P. B. Lilje, A. J. Banday and K. M. Gorski, 2004b, ApJ, 612, 64

\item[] H. K. Eriksen, A. J. Banday, K. M. Gorski and P. B. Lilje, 2004c, ApJ, 612, 633

\item[] H. K. Eriksen et al., 2007a, ApJ, 656, 641

\item[] H. K. Eriksen, A. J. Banday, K. M. Gorski, F. K. Hansen and P. B. Lilje, 2007b, ApJ, 660, 2, L81

\item[] C. Gordon, W. Hu, D. Huterer and T. Crawford, 2005, Phys. Rev. D, 72, 103002

\item[] K. M. Gorski et al., 2005, ApJ, 622, 759

\item[] N. E. Groeneboom, M. Axelsson, D. F. Mota and T. Koivisto, arXiv:1011.5353

\item[] A. Gruppuso et al., 2011, MNRAS, 411, 1445

\item[] V. G. Gurzadyan, A. A. Starobinsky, A. L. Kashin, H. G. Khachatryan and G. Yegorian, 2007, Mod. Phys. Lett. A, 22, 39, 2955

\item[] M. Hansen, A. M. Frejsel, J. Kim, P. Naselsky and F. Nesti, 2011, Phys. Rev. D, 83, 103508

\item[] D. Hanson and A. Lewis, 2009, Phys. Rev. D, 80, 063004

\item[] G. Hinshaw et al., 2003, ApJS, 148, 135

\item[] G. Hinshaw et al., 2007, ApJS, 170, 288

\item[] E. Hivon et al., 2002, ApJ, 567, 2

\item[] K. Ichiki, R. Nagata, J. Yokoyama, 2010, Phys. Rev. D, 81, 083010

\item[] N. Jarosik et al., 2011, ApJS, 192, 14

\item[] J. Kim, P. Naselsky and P. R. Christensen, 2008, Phys. Rev. D, 77, 103002

\item[] J. Kim and P. Naselsky, 2010, ApJ, 714, L265

\item[] A. Kogut et al., 1996, ApJ, 464, L5

\item[] A. Kogut et al., 2003, ApJS, 148, 161

\item[] E. Komatsu et al., 2011, ApJS, 192, 18

\item[] T. S. Koivisto and D. F. Mota, 2011, JHEP, 02, 061

\item[] K. Land and J. Magueijo, 2005a, Phys. Rev. Lett., 95, 071301

\item[] K. Land and J. Magueijo, 2005b, Phys. Rev. D, 72, 101302

\item[] D. Larson et al., 2011, ApJS, 192, 16

\item[] M. Maris, C. Burigana, A. Gruppuso, F. Finelli and J. M. Diego, 2011, MNRAS, 415, 2546

\item[] J. Martin, C. Ringeval, 2004, Phys. Rev. D, 69, 083515

\item[] J. Martin, C. Ringeval, 2006, JCAP, 08, 009

\item[] L. Page et al., 2003, ApJS, 148, 233

\item[] Planck ``Blue Book'', Planck Collaboration 2005, Planck : The Scientific Programme, ESA Publication, ESA-SCI(2005)01

\item[] A. Pontzen and H. V. Peiris, 2010, Phys. Rev. D, 81, 103008

\item[] J. P. Ralston and P. Jain, 2004, Int. J. Mod. Phys. D, 13, 1857

\item[] R. Saha, P. Jain and T. Souradeep, 2006, ApJ, 645, L89

\item[] R. Saha, S. Prunet, P. Jain and T. Souradeep, 2008, Phys. Rev. D, 78, 023003

\item[] P. K. Samal, R. Saha, P. Jain and J. P. Ralston, 2008, MNRAS, 385, 1718

\item[] P. K. Samal et al., 2010, ApJ, 714, 840

\item[] D. J. Schwarz, G. D. Starkman, D. Huterer and C. J. Copi, 2004,  Phys. Rev. Lett., 93, 221301

\item[] D. N. Spergel et al., 2003, ApJS, 148, 175

\item[] M. Tegmark, A. de Oliveira-Costa and A. J. S. Hamilton, 2003, Phys. Rev. D, 68, 123523

\item[] M. S. Turner, 1983, Phys. Rev. D, 28, 1243

\item[] X. Wang, B. Feng, M. Li, X-L. Chen, X. Zhang, 2005, Int. J. Mod. Phys. D, 14, 1347

\end{itemize}

\end{document}